\documentclass{jnmp}
 \usepackage{amsmath}

\JNMPnumberwithin{equation}{section}

\theoremstyle{definition}

\newcommand{\beq}{\begin{equation}}
\newcommand{\eeq}{\end{equation}}
\newcommand{\bdm}{\begin{displaymath}}
\newcommand{\edm}{\end{displaymath}}
\newcommand{\beqr}{\begin{eqnarray}}
\newcommand{\eeqr}{\end{eqnarray}}
\newcommand{\beqrn}{\begin{eqnarray*}}
\newcommand{\eeqrn}{\end{eqnarray*}}
\def\l{\lambda}
\def\R{\mathbb{R}}
\def\Z{\mathbb{Z}}
\def\C{\mathbb{C}}
\def\Oc{\mathbb{O}}
\def\He{\mathbb{H}}
\def\bchi{\boldsymbol{\chi}}

\begin{document}

%
\renewcommand{\evenhead}{J Fern\'andez-N\'u\~nez, W Garc\'{\i}a-Fuertes and A M Perelomov}
\renewcommand{\oddhead}{Irreducible characters and Clebsch-Gordan series for $E_6$}

%
\thispagestyle{empty}

\FirstPageHead{*}{*}{20**}{\pageref{firstpage}--\pageref{lastpage}}{Article}

\copyrightnote{200*}{J Fern\'andez-N\'u\~nez, W Garc\'{\i}a-Fuertes and A M Perelomov}

\Name{Irreducible characters and Clebsch-Gordan series for the exceptional algebra $E_6$: an approach through the quantum Calogero-Sutherland model}

\label{firstpage}

\Author{J Fern\'andez-N\'u\~nez~$^\dag$,  W Garc\'{\i}a-Fuertes~$^{\dag}$ and A M Perelomov~$^\ddag$}

\Address{$^\dag$ Departamento de F\'{\i}sica, Universidad de Oviedo, E-33007 Oviedo, Spain \\
~~E-mail: nonius@uniovi.es (JFN), wifredo@uniovi.es (WGF)\\[10pt]
$^\ddag$ Departamento de F\'{\i}sica Te\'orica, Universidad de Zaragoza, E-50009 Zaragoza, Spain \\
~~E-mail: perelomo@dftuz2.unizar.es}

\Date{Received Month *, 200*; Revised Month *, 200*; 
Accepted Month *, 200*}

\begin{abstract}
\noindent
We re-express the quantum Calogero-Sutherland model for the Lie algebra $E_6$ and the particular value of the coupling constant $\kappa=1$ by using the fundamental irreducible characters of the algebra as dynamical variables. For that, we need to develop a systematic procedure to obtain all the Clebsch-Gordan series required to perform the change of variables. We describe how the resulting quantum Hamiltonian operator can be used to compute more characters and Clebsch-Gordan series for this exceptional algebra.
\end{abstract}

\begin{center}
\it\Large To Francesco Calogero on ocassion of his seventieth birthday.
\end{center}

%
\section{Introduction}
During the three last decades of the past century, a plethora of highly nontrivial mechanical integrable systems 
were discovered, see \cite{ca01, pe90} for comprehensive reviews. Among these, the Calogero-Sutherland  models form a distinguished class. 
The first analysis of a system of this kind was performed by Calogero 
\cite{ca71} who studied, from the quantum standpoint, the dynamics on the 
infinite line of a set of particles interacting pairwise by rational 
plus quadratic potentials, and found that the problem was exactly 
solvable. Soon afterwards, 
Sutherland \cite{su72} arrived to similar results for the quantum problem 
on the circle, this time with trigonometric interaction, 
and Moser \cite{mo75} showed that the classical version of both models 
enjoyed integrability in the Liouville sense. The identification of the 
general scope of these discoveries came with the work 
of Olshanetsky and Perelomov \cite{op76}-\cite{op78}, who realized that it 
was possible to associate models of this kind to all the root systems 
of the simple Lie algebras, and that all these models were integrable, 
both in the classical and in the quantum framework \cite{op81,op83}. 
Nowadays, there is a widespread interest in this type of integrable 
systems, and many mathematical and physical applications for them 
have been found, see for instance \cite{dv00}. 

The study of the form and properties of the Schr\"{o}dinger eigenfunctions for the quantum version of these models constitutes an interesting line of research. In fact, these eigenfunctions have very rich mathematical properties. In particular, for the case with trigonometric potential, if we tune the coupling constants to same especial values, the wave functions correspond to the characters of the simple Lie algebras, while if we select a different tuning, we can make them to coincide with zonal spherical functions. Thus, the Calogero-Sutherland theories provide us with a new tool for computing these quantities. In this spirit, we will describe in the present paper how to use the trigonometric Calogero-Sutherland model to obtain both particular characters and Clebsch-Gordan series for the exceptional Lie algebra $E_6$. The main point of our approach is to express the Calogero-Sutherland Hamiltonian in a suitable set of independent variables, indeed the fundamental characters of $E_6$. The use of such kind of variables has been quite useful to solve the Schr\"{o}dinger equation for the models associated to some classical algebras, \cite{op83}, \cite{pe98a}-\cite{ngp03}.

The organization of the paper is as follows. Section 2. is a reminder of the properties of $E_6$ relevant for the contents of the paper. Section 3. describes the Calogero-Sutherland model associated to $E_6$ and explains how to perform the change of variables mentioned above. Section 4. gives a detailed account of the computation of the Clebsch-Gordan series of $E_6$ needed to pass to the new variables. In Section 5. we present the Hamiltonian in these variables and describe its use for computing new characters and to reduce tensor products of representations. Some conclusions are given in Section 6., and finally, the appendices show some explicit results for characters and Clebsch-Gordan series of $E_6$.

\section{Summary of results on the Lie algebra $E_6$}
In this Section, we review some standard facts about the root and weight systems of the Lie algebra $E_6$, with the aim of fixing the notation and help the reader to follow the rest of the paper. More extensive and sound treatments of these topics can be found in many excellent textbooks, see for instance \cite{ov90}, \cite{otros}. 

The complex Lie algebra $E_6$, the lowest-dimensional one in the $E$-family of exceptional Lie algebras in the Cartan-Killing classification, has dimension 78 and rank 6, as the name suggests. From the geometrical point of view, it admits (with some subtleties, see \cite{baez}) an interpretation which extends the standard-one for the classical algebras: in the same way that these correspond to the isometries of projective spaces over the first three normed division algebras ---$SO(n+1)\simeq {\rm Isom}(\R P^n)$, $SU(n+1)\simeq {\rm Isom}({\C}P^n)$, $Sp(n+1)\simeq {\rm Isom}({\He}P^n)$---, $F_4$, $E_6$, $E_7$ and $E_8$ are the Lie algebras of the projective planes over extensions of the octonions, giving rise to the so-called ``magic square": $F_4\simeq {\rm Isom}({\Oc}P^2)$, $E_6\simeq {\rm Isom}[({\C}\otimes{\Oc})P^2]$,  $E_7\simeq {\rm Isom}[({\He}\otimes{\Oc})P^2]$, $E_8\simeq {\rm Isom}[({\Oc}\otimes{\Oc})P^2]$. In Physics, the most remarkable role played by $E_6$ is in the heterotic ten-dimensional $E_8\times E_8$ superstring theory when the extra six dimensions are compactified to a manifold of $SU(3)$ holonomy: in such a case, one of the $E_8$ breaks to an $E_6$ which gives the Grand Unification group of four-dimensional physics \cite{ramond}.  The Dynkin diagram of $E_6$, see Figure 1, 
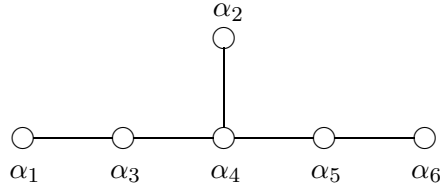
\begin{figure}
\begin{center}
\begin{picture}(80,65)(-50,-8)
\put(0,0){\circle{8}}
\put(4,0){\line(1,0){30}}
\put(38,0){\circle{8}}
\put(42,0){\line(1,0){30}}
\put(76,0){\circle{8}}
\put(0,4){\line(0,1){30}}
\put(0,38){\circle{8}}
\put(-4,0){\line(-1,0){30}}
\put(-38,0){\circle{8}}
\put(-42,0){\line(-1,0){30}}
\put(-76,0){\circle{8}}
\put(-81,-15){$\alpha_1$}
\put(-43,-15){$\alpha_3$}
\put(-5,-15){$\alpha_4$}
\put(33,-15){$\alpha_5$}
\put(71,-15){$\alpha_6$}
\put(-4,46){$\alpha_2$}
\end{picture}
\end{center}
\caption[smallcaption]{The Dynkin diagram for the Lie algebra $E_6$.}
\end{figure}
encodes the euclidean relations among the simple roots, which are
\beqrn
(\alpha_i,\alpha_i)&=&2,\hspace{1.5cm} i=1,2,3,4,5,6\\
(\alpha_4,\alpha_i)&=&-1,\hspace{1.2cm} i=2,3,5\\
(\alpha_1,\alpha_3) &=&(\alpha_5,\alpha_6) =-1,\\
(\alpha_i,\alpha_j)&=&0, \hspace{1.2 cm}{\rm in\ all\ other\ cases}.
\eeqrn
Therefore, the Cartan matrix reads
\bdm
A=\left(\begin{array}{cccccc}
2&0&-1&0&0&0\\
0&2&0&-1&0&0\\
-1&0&2&-1&0&0\\
0&-1&-1&2&-1&0\\
0&0&0&-1&2&-1\\
0&0&0&0&-1&2
\end{array}\right) .
\edm

It is convenient to use a realization of the simple roots in terms of the generating system  $\{\varepsilon_1,\varepsilon_2,\varepsilon_3,\varepsilon_4,\varepsilon_5,\varepsilon_6,\varepsilon\}$ of ${\R}^7$ (endowed with the standard Euclidean metric) satisfying  the conditions $\varepsilon_1+\varepsilon_2+\varepsilon_3+\varepsilon_4+\varepsilon_5+\varepsilon_6=0$, $(\varepsilon_i,\varepsilon_j)=-\frac{1}{6}+\delta_{ij}$, $(\varepsilon,\varepsilon)=\frac{1}{2}$ and $(\varepsilon,\varepsilon_j)=0$ \cite{ov90}. With reference to this system, we have
\begin{eqnarray}
\alpha_1&=&\varepsilon_1-\varepsilon_2,\hspace{1.5cm}\alpha_2=\varepsilon_4+\varepsilon_5+\varepsilon_6+\varepsilon\nonumber\\
\alpha_3&=&\varepsilon_2-\varepsilon_3,\hspace{1.5cm}\alpha_4=\varepsilon_3-\varepsilon_4\label{e61}\\
\alpha_5&=&\varepsilon_4-\varepsilon_5,\hspace{1.5cm}\alpha_6=\varepsilon_5-\varepsilon_6.\nonumber
\end{eqnarray}
The positive roots, which are given by all linear combinations of the forms
\beq
\varepsilon_i-\varepsilon_j,\ \  \varepsilon_i+\varepsilon_j+\varepsilon_k+\varepsilon, \ \ 2\varepsilon,\ \ \ \ i \neq j\neq k,\label{e62}
\eeq
can be classified by heights as indicated in the Table 1. 
\begin{table}[h]
\begin{center}
\begin{tabular}{|c|l|}
\hline
{\rm Height} & {\rm Positive roots} \\
\hline
1 & $\alpha_1,\ \alpha_2,\ \alpha_3,\ \alpha_4,\ \alpha_5,\alpha_6$ \\
\hline
2 & $\alpha_1+\alpha_3,\ \alpha_3+\alpha_4,\ \alpha_4+\alpha_5,\ \alpha_5+\alpha_6,\ \alpha_2+\alpha_4 $\\
\hline
3 & $\alpha_1+\alpha_3+\alpha_4,\ \alpha_3+\alpha_4+\alpha_5,\ \alpha_4+\alpha_5+\alpha_6,\ \alpha_2+\alpha_3+\alpha_4,$\\ & $\  \alpha_2+\alpha_4+\alpha_5$\\
\hline
4 & $\alpha_1+\alpha_3+\alpha_4+\alpha_5,\ \alpha_3+\alpha_4+\alpha_5+\alpha_6,\ \alpha_1+\alpha_2+\alpha_3+\alpha_4,$\\ & $\  \alpha_2+\alpha_3+\alpha_4+\alpha_5,\ \alpha_2+\alpha_4+\alpha_5+\alpha_6$ \\
\hline
5 & $\alpha_1+\alpha_3+\alpha_4+\alpha_5+\alpha_6,\ \alpha_1+\alpha_2+\alpha_3+\alpha_4+\alpha_5,\ \alpha_2+\alpha_3+2\alpha_4+\alpha_5,$\\ &$\ \alpha_2+\alpha_3+\alpha_4+\alpha_5+\alpha_6$ \\
\hline
6 & $ \alpha_1+\alpha_2+\alpha_3+2\alpha_4+\alpha_5,\ \alpha_1+\alpha_2+\alpha_3+\alpha_4+\alpha_5+\alpha_6,$\\ & $\ \alpha_2+\alpha_3+2 \alpha_4+\alpha_5+\alpha_6$\\
\hline
7 & $\alpha_1+\alpha_2+2\alpha_3+2\alpha_4+\alpha_5,\ \alpha_2+\alpha_3+2\alpha_4+2 \alpha_5+\alpha_6,$\\ &$\  \alpha_1+\alpha_2+\alpha_3+2\alpha_4+\alpha_5+\alpha_6$ \\
\hline
8 & $\alpha_1+\alpha_2+2 \alpha_3+2 \alpha_4+\alpha_5+\alpha_6,\ \alpha_1+\alpha_2+\alpha_3+2 \alpha_4+2 \alpha_5+\alpha_6$ \\
\hline
9 & $\alpha_1+\alpha_2+2 \alpha_3+2 \alpha_4+2 \alpha_5+\alpha_6$ \\
\hline
10 & $\alpha_1+ \alpha_2+2 \alpha_3+3 \alpha_4+2 \alpha_5+\alpha_6$\\
\hline
11 & $\alpha_1+ 2\alpha_2+2 \alpha_3+3 \alpha_4+2 \alpha_5+\alpha_6$\\
\hline
\end{tabular}
\caption[smallcaption]{Heights of positive roots.}
\end{center}
\end{table}
The fundamental weights $\lambda_k$ follow from the equation $\alpha_i=\sum_{j=1}^4 A_{ji}\lambda_j$. They are
\beqrn
\lambda_1&=&\varepsilon_1+\varepsilon\\
\lambda_2&=&2\varepsilon\\
\lambda_3&=&\varepsilon_1+\varepsilon_2+2\varepsilon\\
\lambda_4&=&\varepsilon_1+\varepsilon_2+\varepsilon_3+3\varepsilon\\
\lambda_5&=&\varepsilon_1+\varepsilon_2+\varepsilon_3+\varepsilon_4+2\varepsilon\\
\lambda_6&=&\varepsilon_1+\varepsilon_2+\varepsilon_3+\varepsilon_4+\varepsilon_5+\varepsilon .
\eeqrn
The geometry of the weight system is summarized by the relations
\beqrn
(\lambda_i,\lambda_j)=A_{ij}^{-1},
\eeqrn
with  $(A_{ij}^{-1})$ the inverse Cartan matrix.
The Weyl vector is
\bdm
\rho =\frac{1}{2}\sum_{\alpha\in {\cal R}^+}\alpha=\sum_{i=1}^6 \lambda_i=8\alpha_1+11\alpha_2+15\alpha_3+21\alpha_4+15\alpha_5+8\alpha_6
\edm
with ${\cal R}^+$ the set of positive roots of the algebra. The Weyl formula for dimensions applied to the irreducible representation associated to the integral dominant weight $\lambda=m_1\lambda_1+m_2\lambda_2+m_3\lambda_3+m_4\lambda_4+m_5\lambda_5+m_6\lambda_6$ gives
\bdm
\dim R_\lambda=\prod_{\alpha\in {\cal R}^+} \frac{(\alpha, \lambda+\rho)}{(\alpha,\rho)}=\frac{P}{2^5\cdot 3^5\cdot 4^5\cdot 5^4 \cdot 6^3 \cdot 7^3\cdot 8^2 \cdot 9\cdot 10\cdot 11}
\edm
where $P$ is a product extended to the set of positive roots in which the root $\alpha=\sum_{i=1}^6 c_i\,\alpha_i$ contributes with a factor ${\rm ht}(\alpha)+\sum_{i=1}^6 c_i\, m_i$, where ${\rm ht}(\alpha)$ is the height of $\alpha$.
In particular, for the fundamental representations, one finds:
\beqrn
\dim R_{\lambda_1}=27&\ \ \ \ \ \ \ \ \ &\dim R_{\lambda_2}=78\\
\dim R_{\lambda_3}=351&\ \ \ \ \ \ \ \ \ &\dim R_{\lambda_4}=2925\\
\dim R_{\lambda_5}=351&\ \ \ \ \ \ \ \ \ &\dim R_{\lambda_6}=27 .
\eeqrn
Note that, these dimensions reflect the fact, coming from the $\Z_2$ symmetry of the Dynkin diagram, that the representations $R_{\l_1}$ and $R_{\l_6}$ are complex conjugates, and the same is true for $R_{\l_3}$ and $R_{\l_5}$, while $R_{\l_2}$ (the adjoint representation) and $R_{\l_4}$ are real.

\section{The Calogero-Sutherland model associated to the Lie algebra $E_6$}
The Hamiltonian operator for the trigonometric Calogero-Sutherland model related to the root system of a simple Lie algebra has the generic form
\[
H=\frac{1}{2}(p,p)+\sum_{\alpha\in{\cal R}^+}\kappa_\alpha(\kappa_\alpha-1)\sin^{-2}(\alpha,q),
\]
where $q$ and $p$ are vectors with dimensions given by the rank $r$ of the algebra, $(\ ,\ )$ is the usual euclidean inner product in ${\R}^r$, and $\kappa_\alpha=\kappa_\beta$ if $||\alpha||=||\beta||$. In particular, because $E_6$ is simply-laced, the Calogero-Sutherland model associated to $E_6$ depends only on one coupling constant $\kappa$. To write $H$ in a more explicit way, it is convenient to use the orthonormal basis $\{e_i,\ i=1,\dots,6\}$ in $\R^6$. 
The expression of $q$ and $p$ in this basis is simply  $q=\sum_{i=1}^6 q_i\, e_i\ $,  $p=\sum_{i=1}^6 p_i\, e_i$, while
the simple roots are given by:
\beqrn
\alpha_1&=&e_1-e_2\\
\alpha_2&=&\frac{1}{2}\left(-1+\frac{\sqrt{3}}{3}\right)\sum_{j=1}^3 e_j+\frac{1}{2}\left(1+\frac{\sqrt{3}}{3}\right)\sum_{j=4}^6 e_j\\
\alpha_k&=&e_{k-1}-e_k,\ \ \ \ \ k=3,4,5,6.
\eeqrn
The $q$ coordinates are assumed to take values in the $[0,\pi]$ interval, and therefore the Hamiltonian can be interpreted as describing the dynamics of a system of six particles moving on the circle, but notice that there is not translational invariance. We recapitulate some important facts about this model which follow from the general structure of the quantum Calogero-Sutherland models related to Lie algebras \cite{op83}. The ground state energy and (non-normalized) wave function are
\begin{eqnarray*}
E_0(\kappa)&=&2 (\rho, \rho)\kappa^2= 156\kappa^2\nonumber\\
\Psi_0^\kappa(q)&=&{\prod_{\alpha\in {\cal R}^+}\sin^\kappa(\alpha, q)},
\end{eqnarray*}
while the excited states depend on the quantum numbers ${\bf m}=(m_1,m_2,m_3,m_4,m_5,m_6)$, and satisfy
\begin{eqnarray}
H\Psi^\kappa_{\bf m}&=&E_{\bf m}(\kappa)\Psi_{\bf m}^\kappa\nonumber\\
E_{\bf m}(\kappa)&=&2 (\lambda+\kappa\rho,\lambda+\kappa\rho),\label{105}
\end{eqnarray}
where $\lambda$ is the highest weight of the irreducible representation of $E_6$ labelled by ${\bf m}$, i. e. $\lambda=\sum_{i=1}^6 m_i \lambda_i$. By substitution in (\ref{105}) of
\beq
\Psi_{\bf m}^\kappa(q)=\Psi_0^\kappa(q)\Phi_{\bf m}^\kappa(q),\label{107}
\eeq
we are led to the eigenvalue problem
\beq
-\Delta^\kappa\Phi_{\bf m}^\kappa=\varepsilon_{\bf m}(\kappa)\Phi_{\bf m}^\kappa
\label{sch}
\eeq
with
\beq
\Delta^\kappa=\frac{1}{2}\Delta+\kappa\sum_{\alpha\in {\cal R}^+}  {\rm ctg}(\alpha, q)(\alpha,\nabla_q)
\label{4b},
\eeq
and
\beq
\label{energ}
\varepsilon_{\bf m}(\kappa)=E_{\bf m}(\kappa)-E_0(\kappa)= 2(\lambda, \lambda+2\kappa\rho).
\eeq
Taking into account that $A_{jk}^{-1}= (\lambda_j,\lambda_k)$, it is possible to give a more explicit expression for $\varepsilon_{\bf m}(\kappa)$:
\beq
\varepsilon_{\bf m}(\kappa)=2\sum_{j,k=1}^6 A_{jk}^{-1} m_j m_k+4\kappa\sum_{j,k=1}^6 A_{jk}^{-1}m_j.\label{eigenvalues}
\eeq

The main problem is to solve (\ref{sch}). As it has been shown for other algebras \cite{op83}-\cite{pe99}, \cite{ngp03}, the best way to do that is to use a set of independent variables which are invariant under the Weyl symmetry of the Hamiltonian, namely the characters of the six fundamental representations of the algebra $E_6$. Unfortunately, the expression of these characters $z_k$ in terms of the $q$-variables (which play the role of coordinates on the maximal torus of $E_6$) is complicated and makes the direct change of variables from $q_i$ to $z_k$ very cumbersome. We  are forced to follow a much more convenient, indirect route, which has proven be useful for other root systems, \cite{ngp03}. First of all, we can infer from (\ref{4b}) the structure of $\Delta^\kappa$ when written in the $z$-variables:
\beq
\Delta^\kappa=\sum_{j, k=1}^6 a_{jk}(z)\partial_{z_j}\partial_{z_k}+\sum_{j=1}^6 \left[b_j^{(0)}(z)+\kappa b_j^{(1)}(z)\right]\partial_{z_j}
\label{structure}.
\eeq
Now, to obtain the full expressions for the coefficients appearing in (\ref{structure}), we rely on the very fact that makes the Calogero-Sutherland model useful for the purposes of the present paper: for $\kappa=1$, the eigenfunction $\Phi_{\bf m}^{(1)}$ is proportional to the character $\bchi_{\bf m}$ of the irreducible representation of $E_6$ with maximal weight $\sum_{i=1}^6m_i\lambda_i$. This implies that we can compute the combination $b_j(z)=b_j^{(0)}(z)+ b_j^{(1)}(z)$ by simply using that, from (\ref{structure}), $\Delta^{(1)} z_j= b_j(z)$, and thus
\beq
\Delta^{(1)} z_j=b_j(z)=\varepsilon_{\bf m}(1) z_j
\eeq
for $(m_k)=(\delta_{jk})$. Suppose now that we know the expressions in the $z$-variables of all second-order characters, that is, the characters of the form $\bchi_{\l_i+\l_j}$, and we know also the form of the Clebsch-Gordan series for the quadratic products of the fundamental characters, i.e. we know the multiplicities $n_{({\bf m};ij)}$ in 
\beq
z_i z_j=\sum_{({\bf m};ij)} n_{({\bf m};ij)} \bchi_{({\bf m};ij)}
\eeq
for every pair $i,j$. Then, by applying the operator $\Delta^{(1)}$ to the two members of these products we can fix the remaining coefficients  $a_{jk}(z_i)$ through the equations
\[
a_{ij}(z)+a_{ji}(z)+b_i(z) z_j+b_j(z)z_i=\sum_{({\bf m};ij)} c_{({\bf m};ij)} \varepsilon_{({\bf m};ij)}(1)\bchi_{({\bf m};ij)}.
\]
These characters and series are, therefore, all that we need to accomplish the task of fixing the form of the Hamiltonian in the limit $\kappa=1$. Although there are some results already available in the literature \cite{ov90, slan}, a number of the required Clebsch-Gordan series remain, to our knowledge, to be calculated. We have thus developed a systematic strategy, entirely based in a few elementary facts, to obtain them. We devote the next Section to give a description of this strategy.
 
\section{Computation of the quadratic Clebsch-Gordan series}
To compute a particular Clebsch-Gordan series $R_{\l_i}\otimes R_{\l_j}$, we proceed through the following steps:
\begin{enumerate}
\item We elaborate a list of all the irreducible representations which could possibly enter in the series. To this end, starting from the highest weight $\l_i+\l_j$, which is directly given by the characters we are multiplying  $z_i,\,z_j$, we subtract all the integral linear combinations of the simple roots such that the result is an integral dominant weight. To do that we have to express the simple roots in the basis of the fundamental weights, that is to say, the components of the $k$-th fundamental weight are the entries in the $k$-th arrow of the Cartan matrix. It turns out that for the series at stake, the list of the possible representations is never very long, the longest one being the corresponding to the case $z_4^2$ which has 24 terms.

\item We identify some of the representations with nonzero multiplicity by the use of two techniques originally devised by Dynkin \cite{dynkin}: the so-called Dynkin theorem and Dynkin method of parts. We here explain them briefly, and refer the reader to the book by R. N. Cahn \cite{cahn} for a more careful exposition with proofs and examples.
\begin{itemize}
\item Dynkin theorem deals with some series of elements of the root space called chains. A chain is an ordered collection $\{\gamma_1,\gamma_2,\ldots,\gamma_n\}$ such that each element $\gamma_k$ is an integer linear combination of the simple roots which is at right angles with all members of the chain other than $\gamma_{k-1}$ and $\gamma_{k+1}$, but it is not orthogonal to any of these two elements. The theorem establishes that if $\Lambda_1$ and $\Lambda_2$ are integral dominant weights and $\{\Lambda_1,\alpha_{k_1},\alpha_{k_2},\ldots,\alpha_{k_n},\Lambda_2\}$ is a chain in which all the $\alpha_{k_i}$ are simple roots, then $\Lambda_1+\Lambda_2-\sum_{i=1}^n\alpha_{k_i}$ is the highest weight of an irreducible representation entering in the direct product of the representations with highest weights $\Lambda_1$ and $\Lambda_2$. In most cases, the information coming from Dynkin theorem refers only to the second highest weight representation in the product, but sometimes the theorem can be used to get some clues about the multiplicity of other representations beyond that.

\item The method of parts uses the reduction of $E_6$ to several subalgebras, namely those appearing when one of the extreme nodes of the diagram of $E_6$ is removed: $A_5$ for the node corresponding to $\alpha_2$ and two different $D_5$ for the nodes of $\alpha_1$ and $\alpha_6$. Each irreducible representation of $E_6$ contains as a subrepresentation the irreducible representation of these subalgebras which arise by removing the index associated to the node deleted: for example, the representation of $E_6$ with highest weight $\sum_{i=1}^6 m_i \lambda_i$ contains the irreducible representation $m_1\tilde{\lambda}_1+m_3\tilde{\lambda}_2+m_4\tilde{\lambda}_3+m_5\tilde{\lambda}_4+m_6\tilde{\lambda}_5$ of $A_5$, with $\tilde{\lambda}_j$ the fundamental weights of that algebra. Also, the product of two irreducible representations of $E_6$ contains the product of the irreducible subrepresentations of $A_5$ or $D_5$ which make part of the representations being multiplied, and one can take advantage of the fact that the irreducible components of these products of subrepresentations are easily worked out through Young diagrams or by using results available in the literature, see for instance the Reference Chapter in \cite{ov90}. Once these components are identified, they can be converted back into irreducible representations of $E_6$ by reinstating indices in the obvious way: $m_1\tilde{\lambda}_1+m_3\tilde{\lambda}_2+m_4\tilde{\lambda}_3+m_5\tilde{\lambda}_4+m_6\tilde{\lambda}_5$ of $A_5$ gives $\sum_{i=1}^6 m_i \lambda_i$ of $E_6$, but now with an unknown $m_2$. We try to fix this index by simply inspecting the list of possible irreducible representations making part of the product. In many cases there is only one possibility, and thus we conclude that the corresponding representation enters in the product with non-zero multiplicity.
\end{itemize} 
\item We use the orthonormality of the system of irreducible characters, i.e.
\[
\langle\bchi_{\bf m}\,|\,\bchi_{\bf n}\rangle =\int_{E_6}d \mu\,\bchi_{\bf m}^*\, \bchi_{\bf n}=\delta_{\bf m,n},
\]
to fix the multiplicity of some irreducible components, tipically the associated to the fundamental weights. For instance, suppose we want to fix the multiplicity $n_{\l_k}$ of the representation $R_{\lambda_k}$ in the product $z_i z_j$, which is given by $n_{\l_k}=\langle z_k \,|\, z_i z_j\rangle$. Imagine that we have worked out the series $z_k z_i^*$ before of the series $z_i z_j$  . Then, as $\langle z_k \,|\, z_i z_j \rangle=\langle z_k z_i^*\,|\, z_j \rangle$ and $\langle z_k z_i^*\,|\, z_j \rangle$  is nothing else that the multiplicity of $R_{\lambda_j}$ in $z_k z_i^*$, which we know, the problem is solved. Note then that to use orthogonality the order in which we obtain the series is important, and, of course, we should begin by the simplest ones. Note also that in these manipulation we use that, as pointed in Section 2, $z_1^*=z_6$,  $z_3^*=z_5$,   $z_2^*=z_2$ and $z_4^*=z_4$.

\item Once the multiplicities of a number of the irreducible components entering in the product have been fixed by means of the former techniques, we write a Diophantine equation by comparing the dimension of the product with the dimensions of the possible irreducible representations whose multiplicities are yet to be fixed. In most cases, if we have been sufficiently exhaustive in our previous analysis, this Diophantine equation will have only one solution, and then we are done. For a few series, however, we can have to deal with a Diophantine equation with several solutions and, in these cases, to choose the correct one among them, we have to go through one supplementary step.

\item We take advantage of the structure $A_5 \oplus U(1)$ in $E_6$, which is apparent from the expression (\ref{e61}) of the roots of $E_6$ in the generating system $\{\varepsilon_i,\varepsilon\}$: the roots $\alpha_1,\alpha_3,\alpha_4,\alpha_5$ and $\alpha_6$ are given by linear combinations of the $\varepsilon_i$ which are suitable to identify those roots as corresponding to $A_5$, while the root $\alpha_2$ incorporates the new generator $\varepsilon$, which is orthogonal to the others and can be associated with a subalgebra $U(1)$. If we now look to the weights of the fundamental representation of $E_6$, $R_{\lambda_1}$, which are \cite{ov90}
\beq
\varepsilon_i\pm \varepsilon,\ \ \ \ -\varepsilon_i-\varepsilon_j,\label{wfund}
\eeq
we extract the branching structure
\beq
z_1=\tilde{\bchi}_{1,0,0,0,0}\,t+\tilde{\bchi}_{0,0,0,1,0}+\tilde{\bchi}_{1,0,0,0,0}\,t^{-1},
\eeq
where $t$ is the character of $U(1)$ and $\tilde{\bchi}_{m_1,m_2,m_3,m_4,m_5}$ are characters of $A_5$. In the same way, given that the roots (\ref{e62}) of $E_6$ are the weights of the adjoint representation $R_{\lambda_2}$, we have
\beq
z_2=t^2+\tilde{\bchi}_{0,0,1,0,0}\,t+\left(1+\tilde{\bchi}_{1,0,0,0,1}\right)+\tilde{\bchi}_{0,0,1,0,0}\,t^{-1}+t^{-2}.
\eeq
The branching expressions for the remaining fundamental representations follow by taking antisymmetric powers of $R_{\lambda_1}$: $R_{\lambda_3}={\rm Alt}(R_{\lambda_1}\otimes R_{\lambda_1})$, $R_{\lambda_4}={\rm Alt}(R_{\lambda_1}\otimes R_{\lambda_1}\otimes R_{\lambda_1})$, and so on. The results are
\begin{eqnarray}
z_3&=&\tilde{\bchi}_{0,1,0,0,0}\, t^2+\left(\tilde{\bchi}_{1,0,0,1,0}+\tilde{\bchi}_{0,0,0,0,1}\right) \, t+\left(\tilde{\bchi}_{0,1,0,0,0}+\tilde{\bchi}_{2,0,0,0,0}+\tilde{\bchi}_{0,0,1,0,1}\right)\nonumber\\
&+&\left(\tilde{\bchi}_{1,0,0,1,0}+\tilde{\bchi}_{0,0,0,0,1}\right) \, t^{-1}+\tilde{\bchi}_{0,1,0,0,0}\, t^{-2}\nonumber\\
z_4&=&\tilde{\bchi}_{0,0,1,0,0} \, t^3+\left(\tilde{\bchi}_{0,1,0,1,0}+\tilde{\bchi}_{1,0,0,0,1}+1  \right)\, t^2\nonumber\\ &+&\left(\tilde{\bchi}_{1,0,1,0,1}+\tilde{\bchi}_{0,0,0,1,1}+2\tilde{\bchi}_{0,0,1,0,0}+\tilde{\bchi}_{1,1,0,0,0}   \right)\, t\nonumber\\ &+&\left( \tilde{\bchi}_{2,0,0,1,0}+\tilde{\bchi}_{0,1,0,1,0}+2\tilde{\bchi}_{1,0,0,0,1}+\tilde{\bchi}_{0,1,0,0,2}+\tilde{\bchi}_{0,0,2,0,0}+1  \right)\nonumber\\ &+&\left(\tilde{\bchi}_{1,0,1,0,1}+\tilde{\bchi}_{0,0,0,1,1}+2\tilde{\bchi}_{0,0,1,0,0}+\tilde{\bchi}_{1,1,0,0,0}   \right)\, t^{-1}\nonumber\\ &+&\left(\tilde{\bchi}_{0,1,0,1,0}+\tilde{\bchi}_{1,0,0,0,1}+1  \right)\, t^{-2}+\tilde{\bchi}_{0,0,1,0,0} t^{-3}\nonumber\\
z_5&=&\tilde{\bchi}_{0,0,0,1,0}\, t^2+\left(\tilde{\bchi}_{1,0,0,0,0}+\tilde{\bchi}_{0,1,0,0,1}\right) t+\left(\tilde{\bchi}_{0,0,0,1,0}+\tilde{\bchi}_{0,0,0,0,2}+\tilde{\bchi}_{1,0,1,0,0}\right)\nonumber\\&+&\left(\tilde{\bchi}_{1,0,0,0,0}+\tilde{\bchi}_{0,1,0,0,1}\right) \, t^{-1}+\tilde{\bchi}_{0,0,0,1,0}\, t^{-2}\nonumber\\
z_6&=&\tilde{\bchi}_{0,0,0,0,1}\, t+\tilde{\bchi}_{0,1,0,0,0}+\tilde{\bchi}_{0,0,0,0,1}\, t^{-1}.\label{e63}
\end{eqnarray}
Thus, the quadratic products of characters of $E_6$  give some linear combinations of powers of $t$, whose coefficients are sums of irreducible characters of $A_5$ which can be computed from the previous formulas through the usual Young diagrammatic combinatorics. Also, the character of each irreducible component appearing in the product has the same structure, and it can be computed if the expression of the character in terms of the $z$'s is known. In favourable circumstances, by comparing powers of $t$ in both members of the Clebsch-Gordan series one can set some bounds on multiplicities entering in the Diophantine equation, and it can happen that this bound are enough to determine that only one of the solutions is acceptable. 
\end{enumerate} 
As we have seen, when we are computing a series, both in the use of orthogonality relations and in the explotation of the branching rules, we often rely on the form of other series that we should have computed before. Therefore, the order in which the series are obtained is very important. The ordering $z_1^2,z_1 z_2, z_1 z_6, z_1 z_3, z_2^2, z_1 z_4, z_3 z_5, z_2 z_3, z_2 z_4, z_3^2, z_1 z_5, z_3 z_4, z_4^2$ proves to be good enough for a fruitful use of the mentioned techniques \footnote{Note, however, that specially when we need to obtain the expression of one of the second-order characters, it can happen that we have to obtain some cubic series. We can do that with the procedure described, starting always by the character of lowest height among those that we need to calculate.}

Let us now show in a concrete case how all this works. Suppose we want to reduce the product $z_3 z_4$, which corresponds to a representation of dimension $351\times2925=1026675$. We begin by writing a list with all possible dominant weights entering in the series, starting with $\lambda_3+\lambda_4$ and going down in the ordering by height. These weights, along with the dimensions of the corresponding representations, are given in the Table 2. 
\begin{table}[h]
\begin{center}
\begin{tabular}{|l|r|}
\hline
{\rm Representation} & {\rm Dimension} \\
\hline
$R_{\lambda_3+\lambda_4}$&386100 \\
$R_{\lambda_1+\lambda_2+\lambda_5}$&314496\\
$R_{\lambda_1+\lambda_3+\lambda_6}$&112320\\
$R_{2\lambda_5}$&34398\\
$R_{2\lambda_2+\lambda_6}$&46332\\
$R_{\lambda_4+\lambda_6}$&51975\\
$R_{2\lambda_1+\lambda_2}$&19305\\
$R_{\lambda_2+\lambda_3}$&17500\\
$R_{\lambda_1+2\lambda_6}$&7722\\
$R_{\lambda_1+\lambda_5}$&7371\\
$R_{\lambda_2+\lambda_6}$&1728\\
$R_{2 \lambda_1}$&351\\
$R_{\lambda_3}$&351\\
$R_{\lambda_6}$&27\\
\hline
\end{tabular}
\caption[smallcaption]{Representations in $R_{\l_3}\otimes R_{\l_4}$.}
\end{center}
\end{table}
Now, one can see from the metric relations given in Section 2 that $\{ \lambda_3,\alpha_3,\alpha_4,\lambda_4\}$ is a chain, and given that $\lambda_3+\lambda_4-\alpha_3-\alpha_4=\lambda_1+\lambda_2+\lambda_5$, Dynkin theorem guarentees that $R_{\lambda_1+\lambda_2+\lambda_5}$ appears in the series with non-zero multiplicity. Let us next turn to consider the reduction to the subalgebra $A_5$ by deleting the dot corresponding to the root $\alpha_2$ in the Dynkin diagram. This means that the product under consideration can be written $\bchi_{0,\cdot,1,0,0,0} \cdot \bchi_{0,\cdot,0,1,0,0}$ and thus can be related to the product $\tilde{\bchi}_{0,1,0,0,0} \cdot \tilde{\bchi}_{0,0,1,0,0}$ in $A_5$. Then, using Young diagrams we find
\[
\bchi_{0,\cdot,1,0,0,0} \cdot \bchi_{0,\cdot,0,1,0,0}=\bchi_{0,\cdot,1,1,0,0}+\bchi_{1,\cdot,0,0,1,0}+\bchi_{0,\cdot,0,0,0,1}.
\]
Finally, we have to re-introduce the index corresponding to $\lambda_2$, and looking at the table of dominant weights, we see that the first weight in the right-hand member corresponds to $\bchi_{0,0,1,1,0,0}$, while the second can be adjudicated to $\bchi_{1,0,0,0,1,0}$ or $\bchi_{1,1,0,0,1,0}$, and the third to $\bchi_{0,0,0,0,0,1}$ or $\bchi_{0,1,0,0,0,1}$. So, in this case, the reduction to $A_5$ gives quite ambiguous information. Then, we do the reduction to the subalgebra $D_5$ in two possible ways, first by removing the dot corresponding to $\alpha_1$ and then doing the same with the node of $\alpha_6$, and in each case we perform an analysis along the same lines than for $A_5$. This gives us very useful information:
the representations:
\[
R_{\lambda_1+\lambda_3+\lambda_6},\ R_{\lambda_3},\ R_{2\lambda_1+\lambda_2},\ R_{\lambda_1+2\lambda_6},\ R_{\lambda_4+\lambda_6},\ R_{\lambda_2+\lambda_6}
\]
have all non-zero multiplicities. Now, the multiplicity of $R_{\lambda_3+\lambda_4}$ is one because it corresponds to the highest weight in the series. Furthermore, given that $R_{\lambda_1+\lambda_3+\lambda_6}$ has non-zero multiplicity, and taking into account the balance of dimensions, we see that the multiplicity of $R_{\lambda_1+\lambda_2+\lambda_5}$ is necessarily one. 

So far we have used the Dynkin theorem and the method of parts. Let us now try to exploit orthogonality to find out the multiplicity of $R_{\lambda_2+\lambda_3}$ by computing $\langle\bchi_{\lambda_2+\lambda_3}\,|\,z_3 z_4\rangle$. Given that we have been following the order mentioned above, we can by now extract the expression of $\bchi_{0,1,1,0,0,0}$ from the series we had already computed, see (\ref{characters}) below, and we find
\[
\bchi_{0,1,1,0,0,0}=z_2 z_3-z_1 z_5-z_1^2+z_3+z_6,
\]
and thus 
\beqrn
\langle\bchi_{\lambda_2+\lambda_3}\,|\,z_3 z_4\rangle&=&\langle z_2 z_3-z_1 z_5-z_1^2+z_3+z_6 \,|\, z_3 z_4\rangle= \langle z_3 z_5 \,|\,z_2 z_4 \rangle -\langle z_5^2 \,|\,z_4 z_6\rangle\nonumber\\
 &-&\langle z_1 z_5 \,|\,z_4 z_6\rangle +\langle z_3 z_5 \,|\,z_4\rangle+\langle z_5 z_6 \,|\,z_4\rangle =9-6-3+1+1=2.
\eeqrn
We have used that all quadratic products entering in the computation have been computed previously, and given that, all inner products follow from the orthonormality of the irreducible components appearing in each one of them. So, the multiplicity $n_{\lambda_2+\lambda_3}$ of $R_{\lambda_2+\lambda_3}$ is two, and as byproduct of this and of the list of weights obtained by applying the method of parts to $D_5$, we conclude that the multiplicity of $R_{\lambda_1+\lambda_3+\lambda_6}$ is $n_{\lambda_1+\lambda_3+\lambda_6}=1$, otherwise the dimensionality of the right-hand member of the series would exceed that of the left-hand member. Similar use of orthogonality considerations allow us to fix the multiplicities  $n_{\lambda_1+\lambda_5}=2$, $n_{\lambda_2+\lambda_6}=2$, $n_{2 \lambda_1}=1$, $n_{\lambda_3}=1$  and $n_{\lambda_6}=2$. At this point, only five multiplicities remain to be calculated, and we can try to obtain them by solving a Diophantine equation. From the table of dimensions, we write
\[
34398\,n_{2\lambda_5}+46332\,n_{2\lambda_2+\lambda_6}+51975\,n_{\lambda_4+\lambda_6}+19305\,n_{2\lambda_1+\lambda_2}+7722\,n_{\lambda_1+2 \lambda_6}=159732.
\]
From the reduction to $D_5$,  we know that $n_{\lambda_4+\lambda_6}$ and $n_{\lambda_1+2\lambda_6}$ are grater or equal to one, but the other multiplicities could be zero. The equation can be readily see to have three solutions
\begin{eqnarray*}
n_{2\lambda_5}=1,\ \ n_{2\lambda_2+\lambda_6}=0,\ \ n_{\lambda_4+\lambda_6}=1,\ \ n_{2\lambda_1+\lambda_2}=1,\ \ n_{\lambda_1+2\lambda_6}=7,\\
n_{2\lambda_5}=1,\ \ n_{2\lambda_2+\lambda_6}=0,\ \ n_{\lambda_4+\lambda_6}=1,\ \ n_{2\lambda_1+\lambda_2}=3,\ \ n_{\lambda_1+2\lambda_6}=2,\\
n_{2\lambda_5}=1,\ \ n_{2\lambda_2+\lambda_6}=1,\ \ n_{\lambda_4+\lambda_6}=1,\ \ n_{2\lambda_1+\lambda_2}=1,\ \ n_{\lambda_1+2\lambda_6}=1.
\end{eqnarray*}
To fix the correct one, we resort to the branching relations described above. By multiplying the expressions (\ref{e63}) and using the Littlewood-Richardson rule, we find that
\[
z_3 z_4=\sum_{k=-5}^5 a_k\, t^k
\]
with
\beqrn
a_4&=&2\,\tilde{\bchi}_{2,0,0,0,0}+{\rm other\  irreducible\ characters}\\
a_3&=&6\,\tilde{\bchi}_{1,0,0,0,2}+{\rm other\  irreducible\ characters}
\eeqrn
while
\[
\bchi_{2,1,0,0,0,0}=\sum_{k=-4}^4 b_k\, t^k,\ \ \ \ \ \ \ \bchi_{1,0,0,0,0,2}=\sum_{k=-3}^3 c_k\, t^k
\]
with
$b_4=\tilde{\bchi}_{2,0,0,0,0}$ and $ c_3=\tilde{\bchi}_{1,0,0,0,2}$.
Therefore, the multiplicities of $R_{\lambda_1+2\lambda_6}$ and $R_{2\lambda_1+\lambda_2}$ can respectively be non higher than 6 and 2. The only acceptable solution   is then
\[
n_{2\lambda_5}=1,\ \ n_{2\lambda_2+\lambda_6}=1,\ \ n_{\lambda_4+\lambda_6}=1,\ \ n_{2\lambda_1+\lambda_2}=1,\ \ n_{\lambda_1+2\lambda_6}=1
\]
and the series is fixed.

Applying the method that we have just described, the final results we have found for the quadratic Clebsch-Gordan series (expressed here in terms of representations $R_\l$), are
\beqr
\label{series}
 R_{\lambda_1}\otimes R_{\lambda_1}&=& R_{2\lambda_1}\oplus  R_{\lambda_3}\oplus  R_{\lambda_6}, \\ \nonumber 
 R_{\lambda_1}\otimes R_{\lambda_2}&=& R_{\lambda_1+\lambda_2}\oplus  R_{\lambda_5}\oplus  R_{\lambda_1},\\ \nonumber
 R_{\lambda_1}\otimes R_{\lambda_3}&=& R_{\lambda_1+\lambda_3}\oplus  R_{\lambda_4}\oplus  R_{\lambda_1+\lambda_6}\oplus  R_{\lambda_2},\\ \nonumber
 R_{\lambda_1}\otimes R_{\lambda_4}&=& R_{\lambda_1+\lambda_4}\oplus  R_{\lambda_2+\lambda_5}\oplus  R_{\lambda_3+\lambda_6}\oplus  R_{\lambda_1+\lambda_2}\oplus  R_{\lambda_5},\\ \nonumber
 R_{\lambda_1}\otimes R_{\lambda_5}&=& R_{\lambda_1+\lambda_5}\oplus  R_{\lambda_2+\lambda_6}\oplus  R_{\lambda_3}\oplus  R_{\lambda_6},\\ \nonumber
 R_{\lambda_1}\otimes R_{\lambda_6}&=& R_{\lambda_1+\lambda_6}\oplus  
R_{\lambda_2}\oplus  R_{0},\\  \nonumber  
 R_{\lambda_2}\otimes R_{\lambda_2}&=& R_{2\lambda_2}\oplus  R_{\lambda_4}\oplus  R_{\lambda_1+\lambda_6}\oplus  R_{\lambda_2}\oplus  R_{0},\\  \nonumber
 R_{\lambda_2}\otimes R_{\lambda_3}&=& R_{\lambda_2+\lambda_3}\oplus  R_{\lambda_1+\lambda_5}\oplus  R_{\lambda_2+\lambda_6}\oplus  R_{2\lambda_1}\oplus  R_{\lambda_3}\oplus  R_{\lambda_6},\\ \nonumber
 R_{\lambda_2}\otimes R_{\lambda_4}&=& R_{\lambda_2+\lambda_4}\oplus  R_{\lambda_3+\lambda_5}\oplus  R_{\lambda_1+\lambda_2+\lambda_6}\oplus  R_{\lambda_5+\lambda_6}\oplus   R_{\lambda_1+\lambda_3}\oplus  R_{2\lambda_2}\oplus  R_{\lambda_4}\\ \nonumber & &\oplus  R_{\lambda_1+\lambda_6}\oplus  R_{\lambda_2},\\ \nonumber
 R_{\lambda_2}\otimes R_{\lambda_5}&=& R_{\lambda_2+\lambda_5}\oplus  R_{\lambda_3+\lambda_6}\oplus  R_{\lambda_1+\lambda_2}\oplus  R_{2\lambda_6}\oplus  R_{\lambda_5}\oplus  R_{\lambda_1},\\ \nonumber
 R_{\lambda_2}\otimes R_{\lambda_6}&=& R_{\lambda_2+\lambda_6}\oplus  R_{\lambda_3}\oplus  R_{\lambda_6},\\ \nonumber
 R_{\lambda_3}\otimes R_{\lambda_3}&=& R_{2\lambda_3}\oplus  R_{\lambda_1+\lambda_4}\oplus  R_{\lambda_2+\lambda_5}\oplus  R_{2\lambda_1+\lambda_6}\oplus  R_{\lambda_3+\lambda_6}\oplus 2 R_{\lambda_1+\lambda_2}\oplus  R_{2\lambda_6}\\ \nonumber & &\oplus  R_{\lambda_5}\oplus  R_{\lambda_1},\\ \nonumber
 R_{\lambda_3}\otimes R_{\lambda_4}&=& R_{\lambda_3+\lambda_4}\oplus  R_{\lambda_1+\lambda_2+\lambda_5}\oplus  R_{\lambda_1+\lambda_3+\lambda_6}\oplus  R_{2\lambda_5}\oplus  R_{2\lambda_2+\lambda_6}\oplus  R_{\lambda_4+\lambda_6}\oplus  R_{2\lambda_1+\lambda_2}\\ \nonumber
&& \oplus 2 R_{\lambda_2+\lambda_3}\oplus R_{\lambda_1+2\lambda_6}\oplus 2 R_{\lambda_1+\lambda_5}\oplus 2 R_{\lambda_2+\lambda_6}\oplus  R_{2\lambda_1}\oplus  R_{\lambda_3} \oplus  R_{0},\\ \nonumber   
 R_{\lambda_3}\otimes R_{\lambda_5}&=& R_{\lambda_3+\lambda_5}\oplus  R_{\lambda_1+\lambda_2+\lambda_6}\oplus  R_{\lambda_5+\lambda_6}\oplus  R_{\lambda_1+\lambda_3}\oplus  R_{2\lambda_2}\oplus  R_{\lambda_4}\oplus 2 R_{\lambda_1+\lambda_6}\\ \nonumber & &\oplus  R_{\lambda_2}\oplus  R_{0},\\ \nonumber
 R_{\lambda_3}\otimes R_{\lambda_6}&=& R_{\lambda_3+\lambda_6}\oplus  R_{\lambda_1+\lambda_2}\oplus  R_{\lambda_5}\oplus  R_{\lambda_1},\\ \nonumber
 R_{\lambda_4}\otimes R_{\lambda_4}&=& R_{2\lambda_4}\oplus  R_{\lambda_2+\lambda_3+\lambda_5}\oplus  R_{\lambda_1+2\lambda_5}\oplus  R_{2\lambda_3+\lambda_6}\oplus  R_{\lambda_1+2\lambda_2+\lambda_6}\oplus  R_{\lambda_1+\lambda_4+\lambda_6}\\ \nonumber& &\oplus 2 R_{\lambda_2+\lambda_5+\lambda_6}   \oplus 2 R_{\lambda_1+\lambda_2+\lambda_3}\oplus  R_{3\lambda_2}\oplus 2 R_{\lambda_2+\lambda_4}\oplus  R_{2\lambda_1+2\lambda_6}\oplus  R_{\lambda_3+2\lambda_6} \\ \nonumber
&&\oplus  R_{2\lambda_1+\lambda_5}\oplus 3 R_{\lambda_3+\lambda_5}\oplus 4 R_{\lambda_1+\lambda_2+\lambda_6}\oplus  R_{3\lambda_6}\oplus  R_{3\lambda_1}\oplus 2 R_{\lambda_1+\lambda_3}\oplus 2 R_{\lambda_5+\lambda_6}\\ \nonumber
&&\oplus 2 R_{2\lambda_2}\oplus 2 R_{\lambda_4}\oplus 3 R_{\lambda_1+\lambda_6}\oplus  R_{\lambda_2}\oplus  R_{0},\\Ê \nonumber
R_{\lambda_4}\otimes R_{\lambda_5}&=& R_{\lambda_4+\lambda_5}\oplus  R_{\lambda_2+\lambda_3+\lambda_6}\oplus  R_{\lambda_1+\lambda_5+\lambda_6}\oplus   R_{2\lambda_3}\oplus  R_{\lambda_1+2\lambda_2}\oplus  R_{\lambda_1+\lambda_4}\oplus  R_{\lambda_2+2\lambda_6}\\ \nonumber 
&&\oplus 2 R_{\lambda_2+\lambda_5}\oplus  R_{2\lambda_1+\lambda_6}\oplus 2 R_{\lambda_3+\lambda_6}\oplus 2 R_{\lambda_1+\lambda_2}\oplus  R_{2\lambda_6}\oplus  R_{\lambda_5}  \oplus  R_{\lambda_1},\\  \nonumber
R_{\lambda_4}\otimes R_{\lambda_6}&=& R_{\lambda_4+\lambda_6}\oplus  R_{\lambda_2+\lambda_3}\oplus  R_{\lambda_1+\lambda_5}\oplus  R_{\lambda_2+\lambda_6}\oplus  R_{\lambda_3},\\ \nonumber
 R_{\lambda_5}\otimes R_{\lambda_5}&=& R_{2\lambda_5}\oplus  R_{\lambda_4+\lambda_6}\oplus  R_{\lambda_2+\lambda_3}\oplus  R_{\lambda_1+2\lambda_6}\oplus  R_{\lambda_1+\lambda_5}\oplus 2 R_{\lambda_2+\lambda_6}\oplus  R_{2\lambda_1}\\ \nonumber & &\oplus  R_{\lambda_3}\oplus  R_{\lambda_6},\\ \nonumber
R_{\lambda_5}\otimes R_{\lambda_6}&=& R_{\lambda_5+\lambda_6}\oplus  R_{\lambda_4}\oplus  R_{\lambda_1+\lambda_6}\oplus  R_{\lambda_2},\\ \nonumber
R_{\lambda_6}\otimes   R_{\lambda_6}&=& R_{2\lambda_6}\oplus  R_{\lambda_5}\oplus  R_{\lambda_1}.
\eeqr
From these series, the second order characters are
\beqr
\bchi_{1,0,0,0,0,0}&=&z_1,\\  \nonumber
\bchi_{0,1,0,0,0,0}&=&z_2,\\  \nonumber
\bchi_{0,0,1,0,0,0}&=&z_3,\\  \nonumber
\bchi_{0,0,0,1,0,0}&=&z_4,\\  \nonumber
\bchi_{0,0,0,0,1,0}&=&z_5,\\   \nonumber
\bchi_{0,0,0,0,0,1}&=&z_6,\\   \nonumber
\bchi_{2,0,0,0,0,0}&=&z_1^2-z_3-z_6,\\   \nonumber
\bchi_{1,1,0,0,0,0}&=&z_1z_2-z_1-z_5,\\   \nonumber
\bchi_{1,0,1,0,0,0}&=&z_1z_3-z_1z_6-z_4+1,\\  \nonumber
\bchi_{1,0,0,1,0,0}&=&z_1z_4-z_2z_5+z_6^2-z_5,\\  \nonumber
\bchi_{1,0,0,0,1,0}&=&z_1z_5-z_2z_6,\\  \nonumber
\bchi_{1,0,0,0,0,1}&=&z_1z_6-z_2-1,\\   \nonumber
\bchi_{0,2,0,0,0,0}&=&z_2^2-z_4-z_1z_6,\\   \nonumber
\bchi_{0,1,1,0,0,0}&=&z_2z_3-z_1z_5-z_1^2+z_3+z_6,\\   \nonumber
\bchi_{0,1,0,1,0,0}&=&z_2z_4-z_3z_5+z_1z_6-z_2,\\   \nonumber
\bchi_{0,1,0,0,1,0}&=&z_2z_5-z_3z_6-z_6^2+z_5+z_1,\\   \nonumber
\bchi_{0,1,0,0,0,1}&=&z_2z_6-z_6-z_3,\\   \nonumber
\bchi_{0,0,2,0,0,0}&=&z_3^2-z_1z_4-z_1^2z_6+z_3z_6+z_1+z_5,\\   \nonumber
\bchi_{0,0,1,1,0,0}&=&z_3z_4-z_1z_2z_5+z_1z_6^2+z_4z_6-z_6,\\   \nonumber
\bchi_{0,0,1,0,1,0}&=&z_3z_5-z_1z_2z_6+z_1z_6+z_4+z_2-1,\\ \  \nonumber
\bchi_{0,0,1,0,0,1}&=&z_3z_6-z_1z_2,\\   \nonumber
\bchi_{0,0,0,2,0,0}&=&z_4^2-z_2z_3z_5+z_1z_6z_4+z_1^2z_5+z_3z_6^2-2 z_3z_5-z_1z_6-2 z_4+1,\\    \nonumber
\bchi_{0,0,0,1,1,0}&=&z_4z_5-z_2z_3z_6+z_1^2z_6+z_1z_4-z_1,\\   \nonumber
\bchi_{0,0,0,1,0,1}&=&z_4z_6-z_2z_3+z_1^2-z_3,\\  \nonumber
\bchi_{0,0,0,0,2,0}&=&z_5^2-z_4z_6-z_1z_6^2+z_1z_5+z_3+z_6,\\   \nonumber
\bchi_{0,0,0,0,1,1}&=&z_5z_6-z_1z_6-z_4+1,\\  \nonumber
\bchi_{0,0,0,0,0,2}&=&z_6^2-z_5-z_1.
\label{characters}
\eeqr
\section{The Calogero-Sutherland Hamiltonian for $\kappa=1$ and some applications}
After having computed the necessary series and characters, we can now follow the lines indicated towards the end of Section 3 to obtain the Hamiltonian operator for $\kappa=1$. The result for the coeffcients in (\ref{structure}) is
\beqrn
a_ {11}(z)&=&\frac{8}{3}z_1^2 - 4z_3 - 20z_6,  \\ \nonumber
a_ {12} (z)&=&2 z_1 z_2-26 z_1  - 10 z_5,  \\ \nonumber
a_ {13} (z)&=& \frac{10}{3}z_1z_3+18 - 12z_2  - 6z_4 - 18z_1z_6,  \\ \nonumber
a_ {14} (z)&=&  4z_1z_4+ 18z_1 - 10z_1z_2  - 18z_5 - 8z_2z_5 - 8z_3z_6 + 
    8z_6^2,  \\ \nonumber
a_ {15} (z)&=&   \frac{8}{3}z_1z_5 -10 z_3   - 26z_6 - 10z_2z_6,  \\ \nonumber
a_ {16} (z)&=&    \frac{4}{3}z_1z_6 -36  - 12z_2,    \\ \nonumber
a_ {22} (z)&=&    2z_2^2-18  - 6z_2  - 2z_4 - 8z_1z_6,   \\ \nonumber
a_ {23} (z)&=&    4z_2z_3 -24 z_1^2 + 14z_3  - 8z_1z_5 - 2z_6 - 10z_2z_6,  \\ \nonumber
a_ {24} (z)&=&    6z_2z_4-18 z_2 - 12z_2^2 - 10z_1z_3 + 24z_4  - 6z_3z_5 + 26z_1z_6 - 8 z_1z_2z_6 - 10z_5z_6, \\ \nonumber
  a_ {25} (z)&=&   4z_2z_5  -2 z_1 - 10z_1z_2 + 14z_5  - 8z_3z_6 - 24z_6^2 , \\ \nonumber
  a_ {26} (z)&=& 2 z_2 z_6-10 z_3 - 26 z_6,  \\ \nonumber
  a_ {33} (z)&=&   \frac{10}{3}z_3^2+ 14z_1 - 12z_1z_2   - 2z_1 
    z_4 + 16z_5 - 4z_2z_5 - 8z_1^2z_6 + 4z_3z_6 - 6z_6^2 , \\ \nonumber
  a_ {34} (z)&=&   8z_3 z_4+10z_1^2 - 10z_1^2z_2 + 18z_3 - 2z_2z_3  - 6z_1z_2z_5 - 10z_5^2 - 18z_6 + 8z_2z_6 \\ \nonumber
  && - 10z_2^2z_6  -  8z_1z_3z_6 +20z_4z_6 + 8z_1z_6^2,  \\ \nonumber
  a_ {35} (z)&=&    \frac{16}{3}z_3 z_5  -36  + 24z_2 - 12z_2^2 - 10z_1z_3 + 24z_4  - 16z_1z_6 - 8z_1z_2z_6 - 10z_5z_6,  \\ \nonumber
  a_ {36} (z)&=&     \frac{8}{3}z_3z_6-26 z_1 - 10z_1z_2 - 10z_5 ,   \\ \nonumber
  a_ {44} (z)&=&    6 z_4^2 -4 z_1^3 - 6z_2^3 + 18z_1z_3 - 6z_1z_2z_3 - 18z_4 + 18z_2z_4  + 8z_1^2z_5 - 18z_3z_5 \\ \nonumber
   &&- 2z_2z_3z_5 -  4z_1z_5^2 -18z_1z_6 + 14z_1z_2z_6 - 4z_1z_2^2z_6 - 4z_3^2z_6 + 8z_1z_4z_6 \\ \nonumber & &+ 18z_5z_6 - 6z_2z_5z_6 + 8z_3z_6^2 - 4z_6^3,  \\ \nonumber
    a_ {45} (z)&=&   8z_4z_5  -18 z_1 + 8 z_1z_2 - 10z_1z_2^2 - 10z_3^2 + 20z_1z_4 + 18z_5 - 2z_2z_5+ 8z_1^2z_6 \\Ê\nonumber
&&- 6z_2z_3z_6  -  8z_1 z_5z_6 +10z_6^2 - 10z_2z_6^2,  \\ \nonumber
  a_ {46} (z)&=&  4 z_4z_6 +8z_1^2 - 18z_3 - 8z_2z_3 - 8z_1z_5 + 18z_6 - 10z_2z_6,    \\ \nonumber
  a_ {55} (z)&=& \frac{10}{3}z_5^2   -6 z_1^2 + 16z_3 - 4z_2z_3 + 4z_1z_5 +   14z_6 - 12z_2z_6 - 2z_4z_6 - 8z_1z_6^2,  \\ \nonumber
  a_ {56} (z)&=&   \frac{10}{3}z_5z_6 +18 - 12z_2 - 6z_4 - 18z_1z_6,    \\ \nonumber
  a_ {66} (z)&=&   \frac{4}{3}z_6^2  -10 z_1 - 2z_5,   \\
 b_1(z)&=&\frac{104}{3}z_1,\ 
 b_2(z)= 48z_2,\ b_3(z)=\frac{200}{3}z_3,\\ \nonumber  
 b_4(z)&=&96z_4,\  b_5(z)=\frac{200}{3}z_5,\  b_6(z)=\frac{104}{3}z_6.\nonumber
\eeqrn
With the explicit expression at our disposal, we can now try to use the Schr\"{o}dinger equation as an efficient mean to compute particular characters of $E_6$. Given that all these characters are polynomials in the $z$ variables, the Schr\"{o}dinger equation can be solved by applying a systematic procedure, which is suitable to be implemented in a computer program able to carry out symbolic calculations. We propose two alternative methods to find the Schr\"{o}dinger eigenfunctions:
\begin{enumerate}
\item Given a weight $n_1\lambda_1+n_2\lambda_2+n_3\lambda_3+n_4\lambda_4+n_5\lambda_5+n_6\lambda_6$, let us denote $z^{\bf n}=z_1^{n_1}z_2^{n_2}z_3^{n_3}z_4^{n_4}z_5^{n_5}z_6^{n_6}$. The operator $\Delta^{(1)}$ acting on $z^{\bf n}$ gives 
\beq
\Delta^{(1)} z^{\bf n}=\sum_{\beta\in\Lambda} k_{\beta,{\bf n}}\, z^{\bf n-\beta}\label{e64}
\eeq
where  $\Lambda$ only includes integral linear combinations of the simple roots with non-negative coefficients   and, of course, in the exponent of (\ref{e64}) we express $\beta$ in the basis of fundamental weights. In particular, $k_{0,{\bf n}}=\varepsilon_{\bf n}(1)$. The eigenfunctions $\bchi_{\bf m}$ can be written as
\[
\bchi_{\bf m}=\sum_{\mu \in Q^+({\bf m})} c_\mu z^{\bf m-\mu}, \ \ \ \ c_0=1,
\]
where again the $\mu$ in $Q^+({\bf m})$ are integral linear combinations of the simple roots with non-negative coefficients such  that  they do not give rise to negative powers of the $z$'s. By substituting in the Schr\"{o}dinger equation we find the iterative formula
\[
c_\mu=\frac{1}{\varepsilon_{\bf m}(1)-\varepsilon_{\bf m-(\mu-\beta)}(1)}\sum_{\beta\in\Lambda,\beta\neq 0}k_{\beta-{\bf m}-(\mu-\beta)}\, c_{\mu-\beta}.
\]
To use this formula in practice, one should take into account the heights of the $\mu's$ involved, because each coefficient $c_\mu$ can depend only on some of the $c_\nu$ such that ${\rm ht}(\nu)<{\rm ht}(\mu)$.

\item The Clebsch-Gordan series for the product $z_1^{m_1}z_2^{m_2}z_3^{m_3}z_4^{m_4}z_5^{m_5}z_6^{m_6}$ reads
\[
z_1^{m_1}z_2^{m_2}z_3^{m_3}z_4^{m_4}z_5^{m_5}z_6^{m_6}=\bchi_{\bf m}+\sum_{\beta\in R_{\bf m}}n_{\beta} \bchi_{\bf m-\beta}.
\]
Here it is not difficult, in each particular case, to elaborate a list with all the elements in $R_{\bf m}$ (i.e., the integral dominant weights appearing in the series). Furthermore, the operator $\Delta^{(1)}-\varepsilon_{\bf n}(1)$ annihilates the character $\bchi_{\bf n}$. Having this into account, we can make use of the simple-looking formula
\[
\bchi_{\bf m}=\Big\{\prod_{\beta \in R_{\bf m}}\left(\Delta^{(1)}-\varepsilon_{\bf m-\beta}(1)\right)\Big\} z^{\bf m}
\]
to obtain the eigenfunctions.
\end{enumerate}
Through any of these methods, it is possible to compute the characters rather quickly. As an illustration, we offer a list of the third order characters in the Appendix A.

Once we have a method for the computation of the characters, we can extend it to produce an algorithm for calculating the Clebsch-Gordan series. Suppose that we want to obtain the series for $\bchi_{\bf m}\cdot\bchi_{\bf n}$. We  list the possible dominant weights entering in the series arranged by heights
\[
\bchi_{\bf m}\cdot\bchi_{\bf n}=\bchi_{\bf m+n}+n_{\mu_1} \bchi_{\mu_1}+n_{\mu_2} \bchi_{\mu_2}+\ldots
\]
The multiplicity  $n_{\mu_1}$ is simply the difference between the coefficients of $z^{\bf \mu_1}$ in $\bchi_{\bf m}\cdot\bchi_{\bf n}$ and in $\bchi_{\bf m+n}$. Then, $n_{\mu_2}$ is the difference between the coefficient of $z^{\bf \mu_2}$ in $\bchi_{\bf m}\cdot\bchi_{\bf n}$ and the sum of the corresponding coefficients in $\bchi_{\bf m+n}$ and $\bchi_{\bf \mu_1}$, and so on. As an example, we present in Appendix B a list with all the cubic Clebsch-Gordan series.

The approach we are describing is also useful to find the general structure of the series for product of some specific types. Let us consider, for instance, series of the type $z_1 \bchi_{n\lambda_1}$ with arbitrary $n$. If we express the weights of the representation $R_{\lambda_1}$ (\ref{wfund}) in the basis of fundamental weights, we see that there are only three whose coefficients for $\lambda_i$, $i\neq 1$, are all non-negative: $\lambda_1, -\lambda_1+\lambda_3$ and $-\lambda_1+\lambda_6$, hence, the form of the series should be
\beq
z_1 \bchi_{n,0,0,0,0,0}=\bchi_{n+1,0,0,0,0,0}+ a \bchi_{n-1,0,1,0,0,0}+b \bchi_{n-1,0,0,0,0,1}, \label{ser11}
\eeq
where we have to fix $a$ and $b$. Now, by solving the Schr\"{o}dinger equation by means of the first of the two methods described above, one finds
\begin{eqnarray*}
\bchi_{n,0,0,0,0,0}&=&z_1^{n}+(1-n) z_1^{n-2} z_3-z_1^{n-2}z_6+\ldots\\
\bchi_{n-1,0,1,0,0,0}&=&z_1^{n-1} z_3-z_1 z_6+\ldots\\
\bchi_{n-1,0,0,0,0,1}&=&z_1^{n-1} z_6+\ldots.
\end{eqnarray*}
If we substitute in (\ref{ser11}), we can solve for $a$ and $b$, $a=b=1$. We can now check that with these coefficients, the balance of dimensions in (\ref{ser11}) is correct.

We list below the series of the form $z_1 \bchi_{n\lambda_k}$ obtained through the same procedure. Note that the series $z_6 \bchi_{n\lambda_j}$ immediately follow by duality.
\begin{eqnarray*}
z_1 \bchi_{0,n,0,0,0,0}&=&\bchi_{1,n,0,0,0,0}+  \bchi_{0,n-1,0,0,1,0}+\bchi_{1,n-1,0,0,0,0}\\
z_1 \bchi_{0,0,n,0,0,0}&=&\bchi_{1,0,n,0,0,0}+  \bchi_{0,0,n-1,1,0,0}+ \bchi_{0,1,n-1,0,0,0}+\bchi_{1,0,n-1,0,0,1}\\
z_1 \bchi_{0,0,0,n,0,0}&=&\bchi_{1,0,0,n,0,0}+ \bchi_{0,1,0,n-1,1,0}+ \bchi_{0,0,1,n-1,0,1}+\bchi_{1,1,0,n-1,0,0}+  \bchi_{0,0,0,n-1,1,0}\\
z_1 \bchi_{0,0,0,0,n,0}&=&\bchi_{1,0,0,0,n,0}+ \bchi_{0,1,0,0,n-1,1}+\bchi_{0,0,1,0,n-1,0}+\bchi_{0,0,0,0,n-1,1}\\
z_1 \bchi_{0,0,0,0,0,n}&=&\bchi_{1,0,0,0,0,n}+ \bchi_{0,1,0,0,0,n-1}+\bchi_{0,0,0,0,0,n-1}.
\end{eqnarray*}
\section{Conclusions}
In this paper we have shown how the Calogero-Sutherland Hamiltonian for the Lie algebra $E_6$ can be used to compute both Clebsch-Gordan series and characters of that algebra. The treatment we have presented can be applied to the cases of other simple algebras. It can be also extended to deal with the system of orthogonal polynomials based on $E_6$ for general values of the parameter $\kappa$. This way in which this should be done is the subject of a research now in progress and will be published elsewhere.
\section*{Acknowledgements} 
This work has been partially supported by the spanish 
Ministerio de Educaci\'{o}n y Ciencia under grants BFM2003-02532 (J.F.N) and BFM2003-00936 / FISI (W.G.F and A.M.P).
\section*{Appendix A: A list of the characters of $E_6$ of third order.}
\beqrn
\bchi_{000300} &=& -1 + z_2 + z_1 z_2 z_3 + z_3^3 + 3 z_4 - 2 z_2 z_4 -z_1 z_2 z_3 z_4 - 
    3 z_4^2 + z_2 z_4^2 + z_4^3 \\ &+& z_1^2 z_2^2 z_5+  2 z_3 z_5 - z_2^2 z_3 z_5 + 
    z_1 z_3^2 z_5 - 2 z_3 z_4 z_5- 2 z_2 z_3 z_4 z_5 + z_1 z_5^2 + z_1 z_2^2 z_5^2 \\ &+& 
    z_3^2 z_5^2 - z_1 z_4 z_5^2 + z_5^3 + 3 z_1 z_6 - 2 z_1 z_2 z_6 + z_1 z_2^2 z_6 - z_1^2 z_2 z_3 z_6 + z_3^2 z_6+ z_2^2 z_3^2 z_6 \\ &-& 6 z_1 z_4 z_6 + 2 z_1 z_2 z_4 z_6 - 
    z_1 z_2^2 z_4 z_6 - z_3^2 z_4 z_6 + 3 z_1 z_4^2 z_6 + z_2 z_5 z_6 - 2 z_1 z_3 z_5 z_6 \\ &-& 
    z_1 z_2 z_3 z_5 z_6- z_2 z_4 z_5 z_6 - z_1^2 z_5^2 z_6 + z_3 z_5^2 z_6 - 3 z_1^2 z_6^2 + 
    z_1^2 z_2 z_6^2 - z_1^2 z_2^2 z_6^2 \\ &+& z_2^2 z_3 z_6^2 - z_1 z_3^2 z_6^2 +    3 z_1^2 z_4 z_6^2- z_1 z_2 z_5 z_6^2 + z_1^3 z_6^3\\ 
\bchi_{001110} &=& z_1^3 - z_1^3 z_2 + z_2^2 - z_1 z_3 + z_1 z_2 z_3 + z_1 z_2^2 z_3 - z_4 - 
    z_2^2 z_4 + z_4^2 + z_1^2 z_5 - z_1^2 z_2 z_5 \\ &-& z_3 z_5 + z_3 z_4 z_5- z_1 z_2 z_5^2 - 
    z_1 z_6 + z_1 z_2 z_6 - z_1 z_2^2 z_6 - z_2 z_3^2 z_6 + z_1 z_4 z_6 - z_5 z_6 \\ &+& 
    z_2 z_5 z_6 + z_2^2 z_5 z_6 + z_1 z_3 z_5 z_6- z_1^2 z_6^2 + z_1^2 z_2 z_6^2 + z_3 z_6^2 - z_2 z_3 z_6^2 + z_6^3 - z_2 z_6^3\\ 
  \bchi_{001200} &=& -z_1^2 + z_1^2 z_2 + 2 z_3 + z_1 z_3^2 + z_1^2 z_4 - z_1^2 z_2 z_4 - 3 z_3 z_4 + z_3 z_4^2 + z_1 z_2 z_5 + z_1 z_2^2 z_5 \\ &+& 2 z_1^2 z_3 z_5- 2 z_3^2 z_5 - 
    z_2 z_3^2 z_5 - z_1 z_4 z_5 - z_1 z_2 z_4 z_5 + z_2 z_5^2 + z_2^2 z_5^2 + z_1 z_3 z_5^2 \\ &-&
    z_4 z_5^2 + 2 z_6 + z_1^3 z_6 - z_2 z_6 - z_1^3 z_2 z_6+ z_2^2 z_6 - 2 z_1 z_3 z_6 + z_1 z_2^2 z_3 z_6 - 4 z_4 z_6 \\ &+& z_2 z_4 z_6 - z_2^2 z_4 z_6 + 2 z_4^2 z_6 - 
    z_1^2 z_5 z_6 - 2 z_3 z_5 z_6 - z_2 z_3 z_5 z_6- z_1 z_5^2 z_6 - 3 z_1 z_6^2 \\ &+& z_1 z_2 z_6^2 - z_1 z_2^2 z_6^2 - z_1^2 z_3 z_6^2 + z_3^2 z_6^2 + 3 z_1 z_4 z_6^2 +  z_5 z_6^2 - z_2 z_5 z_6^2+ z_1^2 z_6^3 + z_3 z_6^3\\ 
\bchi_{002001} &=& z_1^3 - z_2 - 2 z_1 z_3 - z_1 z_2 z_3 + z_4 + z_2 z_4 + z_1^2 z_5 - z_3 z_5 + 
    z_1 z_2 z_6 + z_3^2 z_6 \\ &-& z_1 z_4 z_6 - z_1^2 z_6^2 + z_3 z_6^2\\ 
\bchi_{002010} &=& -z_1^2 z_2 + z_1^2 z_2^2 + z_2 z_3 - z_2^2 z_3 - z_1^2 z_4 + 2 z_3 z_4 + 
    z_1 z_5 - z_1 z_2 z_5 + z_3^2 z_5 \\ &-& z_1 z_4 z_5 + z_5^2 - z_2^2 z_6- z_1 z_2 z_3 z_6 + 
    2 z_4 z_6 + z_2 z_4 z_6 + z_1 z_2 z_6^2 - z_5 z_6^2\\ 
\bchi_{002100} &=& z_1 z_2 + z_1^2 z_3 - z_3^2 + z_1 z_4 - z_1 z_2 z_4 + z_3^2 z_4 - z_1 z_4^2 + 
    z_1^3 z_5 + z_2^2 z_5 - z_1 z_3 z_5 \\ &-& z_1 z_2 z_3 z_5 + z_4 z_5 + z_2 z_4 z_5 + z_1^2 z_5^2 - z_3 z_5^2 + z_1^2 z_6 - 2 z_1^2 z_2 z_6 + z_1^2 z_2^2 z_6 - 2 z_3 z_6\\ &+&z_2 z_3 z_6 - z_2^2 z_3 z_6 - 2 z_1^2 z_4 z_6 + 3 z_3 z_4 z_6- z_1 z_5 z_6 - z_5^2 z_6 -  z_6^2 - z_1^3 z_6^2 + z_2 z_6^2 \\ &-&z_2^2 z_6^2 + 2 z_1 z_3 z_6^2 + z_4 z_6^2 + z_1 z_6^3 \\ 
 \bchi_{003000} &=& z_1^3 z_2 + z_1 z_3 - 2 z_1 z_2 z_3 + z_3^3 - z_4 + z_2 z_4 - 2 z_1 z_3 z_4 + 
    z_4^2 + z_1^2 z_2 z_5 + z_3 z_5 \\ &-& z_2 z_3 z_5 - z_1 z_2 z_6- 2 z_1^2 z_3 z_6 + 2 z_3^2 z_6 + z_1 z_4 z_6 + z_5 z_6 - z_2 z_5 z_6 + z_3 z_6^2\\ 
\bchi_{010200} &=& 
   1 + z_2 + z_1 z_3 - 2 z_4 - 2 z_2 z_4 - z_1 z_3 z_4 + z_4^2 + z_2 z_4^2 + 
    2 z_1^2 z_2 z_5 - z_3 z_5 \\ &-& 3 z_2 z_3 z_5 - z_2^2 z_3 z_5 - z_3 z_4 z_5 + z_1 z_2 z_5^2 - 
    2 z_1 z_6 - z_1^2 z_3 z_6 + z_2 z_3^2 z_6 + 2 z_1 z_4 z_6 \\ &+&z_5 z_6 - z_4 z_5 z_6+z_1^2 z_6^2 - z_1^2 z_2 z_6^2+ 2 z_2 z_3 z_6^2 - z_1 z_5 z_6^2\\ 
\bchi_{011010} &=& 
   -z_1^3 + z_1 z_3 + z_1 z_2 z_3 - 2 z_1^2 z_5 + 2 z_3 z_5 + z_2 z_3 z_5 - z_1 z_5^2 - z_1 z_6 + z_1 z_2 z_6 \\ &-& z_1 z_2^2 z_6 - z_3^2 z_6 + z_1 z_4 z_6 + z_5 z_6+ z_2 z_5 z_6 + 
    2 z_1^2 z_6^2 - 2 z_3 z_6^2 - z_6^3\\ 
\bchi_{011100} &=& z_1^2 - z_2 z_3 - z_1^2 z_4 + z_2 z_3 z_4 + z_1 z_2 z_5 - z_1 z_2^2 z_5 - 
    z_3^2 z_5 + z_2 z_5^2 - z_1^3 z_6 + z_1 z_3 z_6 \\ &+& z_1 z_2 z_3 z_6 +z_1^2 z_5 z_6 - z_3 z_5 z_6 - z_5 z_6^2\\ 
\bchi_{012000} &=& 
   -z_1^2 z_3 + z_3^2 + z_2 z_3^2 - z_1 z_2 z_4 + z_2 z_5 - z_1 z_3 z_5 + z_4 z_5 + z_3 z_6 + 
    z_1 z_5 z_6 - z_2 z_6^2\\ 
\bchi_{020100} &=& 
   -z_2^2 - z_1 z_3 + z_4 + z_2^2 z_4 - z_4^2 + z_1^2 z_5 - z_3 z_5 - z_2 z_3 z_5 + 
    z_1 z_5^2 + z_1 z_6 + z_1 z_2 z_6 \\ &+& z_3^2 z_6 - 2 z_1 z_4 z_6 - z_5 z_6 - z_1^2 z_6^2 + 
    z_3 z_6^2\\
\bchi_{021000} &=& 
   -z_1^2 z_2 + z_2 z_3 + z_2^2 z_3 - z_3 z_4 + z_1 z_5 - z_1 z_2 z_5 + z_5^2 - z_4 z_6\\ 
\bchi_{030000} &=& z_2^3 + z_1 z_3 - z_4 - 2 z_2 z_4 + z_3 z_5 - 2 z_1 z_2 z_6 + z_5 z_6\\ 
\eeqrn
\beqrn
\bchi_{100101} &=& z_1^3 - z_1 z_3 - z_1 z_2 z_3 + z_3 z_5 - 2 z_1 z_6 + z_1 z_2 z_6 + z_1 z_4 z_6 - z_5 z_6 - z_2 z_5 z_6 + z_6^3\\ 
\bchi_{100110} &=& -z_1^2 z_2 + z_2 z_3 + z_2^2 z_3 + z_1^2 z_4 - z_3 z_4 - z_1 z_2 z_5 + 
    z_1 z_4 z_5 - z_5^2 - z_2 z_5^2 + z_6 \\&+& z_1^3 z_6 - z_1 z_3 z_6 - z_1 z_2 z_3 z_6 - 
    z_4 z_6 + z_3 z_5 z_6 - 2 z_1 z_6^2 + z_1 z_2 z_6^2 + z_5 z_6^2\\ 
\bchi_{100200} &=& z_1 z_2 + z_3^2 - z_1 z_4 - z_1 z_2 z_4 + z_1 z_4^2 + z_1^3 z_5 + z_2 z_5 - 
    z_1 z_3 z_5 - z_1 z_2 z_3 z_5 \\ &-& z_2 z_4 z_5 + z_3 z_5^2 - z_1^2 z_2 z_6 + z_2^2 z_3 z_6 + 
    z_1^2 z_4 z_6 - z_3 z_4 z_6 - z_1 z_5 z_6\\ 
\bchi_{101001} &=& z_1^2 - z_1^2 z_2 - z_3 + z_1 z_5 + z_2 z_6 + z_1 z_3 z_6 - z_4 z_6 - z_1 z_6^2\\ 
\bchi_{101010} &=& -z_1 + z_1 z_2^2 - z_2 z_5 + z_1 z_3 z_5 - z_4 z_5 + z_1^2 z_6 - 
    z_1^2 z_2 z_6 - z_3 z_6 - z_6^2 + z_2 z_6^2\\ 
\bchi_{101100} &=& z_2 + z_1 z_3 + z_4 - z_2 z_4 + z_1 z_3 z_4 - z_4^2 + z_1^2 z_5 -
    z_1^2 z_2 z_5 + z_1 z_5^2 - 2 z_1 z_2 z_6 \\ &+& z_1 z_2^2 z_6- z_1 z_4 z_6 - z_5 z_6\\ 
\bchi_{102000} &=& z_1^2 + z_1^2 z_2 - z_2 z_3 + z_1 z_3^2 - z_1^2 z_4 - z_3 z_4 + z_1 z_5 + 
    z_1 z_2 z_5 - z_1^3 z_6 - z_2 z_6\\ 
\bchi_{110001} &=& -z_2^2 - z_1 z_3 + z_4 + z_1 z_2 z_6 - z_5 z_6\\ 
\bchi_{110010} &=& z_2 z_3 - z_1 z_5 + z_1 z_2 z_5 - z_5^2 - z_6 + z_2 z_6 - z_2^2 z_6 - z_1 z_3 z_6 + z_4 z_6 + z_1 z_6^2\\ 
\bchi_{110100} &=& z_1 - z_1 z_2 - z_1 z_4 + z_1 z_2 z_4 - z_2 z_5 - z_2^2 z_5 - z_1 z_3 z_5 + 
    z_2 z_3 z_6 + z_1 z_5 z_6 \\ &-& z_6^2+ z_2 z_6^2\\ 
\bchi_{111000} &=& -z_1^3 + z_1 z_3 + z_1 z_2 z_3 - z_4 - z_2 z_4 - z_1^2 z_5 + z_1 z_6 + z_5 z_6\\ 
\bchi_{120000} &=& -z_1 z_2 + z_1 z_2^2 - z_1 z_4 - z_2 z_5 - z_1^2 z_6 + z_3 z_6 + z_6^2\\ 
\bchi_{200001} &=& -z_1 z_2 + z_5 + z_1^2 z_6 - z_3 z_6 - z_6^2\\ 
\bchi_{200010} &=& z_2^2 - z_4 + z_1^2 z_5 - z_3 z_5 - z_1 z_2 z_6\\ 
\bchi_{200100} &=& z_3 + z_1^2 z_4 - z_3 z_4 - z_1 z_2 z_5 + z_5^2 - z_2 z_6 + z_2^2 z_6 - 2 z_4 z_6\\ 
\bchi_{201000} &=& 
   z_1 + z_1 z_2 + z_1^2 z_3 - z_3^2 - z_1 z_4 + z_2 z_5 - z_1^2 z_6 - z_3 z_6\\ 
\bchi_{210000} &=& -z_1^2 + z_1^2 z_2 - z_2 z_3 - z_1 z_5 + z_6\\ 
\bchi_{300000} &=& z_1^3 + z_2 - 2 z_1 z_3 + z_4 - z_1 z_6\\ 
\end{eqnarray*}

\section*{Appendix B: A list of cubic Clebsch-Gordan series for $E_6$.}
\begin{eqnarray*}
z_4^3&=&  \bchi_{000300} + 2   \bchi_{011110} + \bchi_{002020} + \bchi_{120020} + \bchi_{022001} + 3   \bchi_{002101} \\ &+& 3   \bchi_{120101}+ 3   \bchi_{100120} + 2   \bchi_{013000} + 2   \bchi_{030011} + 3   \bchi_{100201} + 8   \bchi_{111011} \\ &+& 2   \bchi_{010030} + 10   \bchi_{010111} + 3   \bchi_{220002} + 3   \bchi_{102002} + 2   \bchi_{131000} + 3   \bchi_{200021} \\ &+& 9   \bchi_{001021} + 7   \bchi_{200102} + 10   \bchi_{111100}+ 9   \bchi_{021002} + 4   \bchi_{030100} + 8   \bchi_{010200} \\ &+& 9   \bchi_{220010} + 12   \bchi_{001102} + 9   \bchi_{102010}+ 12   \bchi_{200110} + 21   \bchi_{021010} + 18   \bchi_{110012} \\ &+& 16   \bchi_{130001} + 6   \bchi_{020003} + \bchi_{300003} + 18   \bchi_{211001} + 5   \bchi_{040000} + 27   \bchi_{001110} \\ &+&30   \bchi_{110020} + 6   \bchi_{202000} + 6   \bchi_{000022} +30   \bchi_{012001} + 62   \bchi_{110101} + 10   \bchi_{101003} \\ &+&13   \bchi_{000103} + 10   \bchi_{000030} + 42   \bchi_{020011} + 10   \bchi_{300011} + 4   \bchi_{100004} + 62   \bchi_{101011} \\ &+&10   \bchi_{003000} + 6   \bchi_{320000}+ 42   \bchi_{121000} + 13   \bchi_{300100} + 42   \bchi_{210002} + 58   \bchi_{000111} \\ &+& 73   \bchi_{011002} + 58   \bchi_{101100} + 73   \bchi_{210010} + 42   \bchi_{020100} + 4   \bchi_{400001} + 39   \bchi_{000200} \\ &+& 117   \bchi_{011010} + 98   \bchi_{120001} + 57   \bchi_{100012} + 28   \bchi_{010003} + 65   \bchi_{100020} + 57   \bchi_{201001} \\ &+& 65   \bchi_{002001} + 28   \bchi_{310000} + 156   \bchi_{100101} + 119   \bchi_{010011} + 25   \bchi_{030000} + 51   \bchi_{200002}
\eeqrn
\beqrn
 &+&119   \bchi_{111000} + 87   \bchi_{200010} + 106   \bchi_{010100} + 87   \bchi_{001002} + 128   \bchi_{001010} + 150   \bchi_{110001} 
\\ &+& 16   \bchi_{000003} + 16   \bchi_{300000}+ 72   \bchi_{101000} + 72   \bchi_{000011} + 39   \bchi_{020000}+76   \bchi_{000100} \\ &+& 50   \bchi_{100001} + 21   \bchi_{010000} + 2   \bchi_{000000}\\
z_3 z_4 z_5&=&  \bchi_{001110} + \bchi_{110020} + \bchi_{012001} + \bchi_{003000} + 2   \bchi_{110101} + 3   \bchi_{101011} +\bchi_{000030}\\ &+& 2   \bchi_{020011} + 2   \bchi_{121000} + 4   \bchi_{000111} + 2   \bchi_{210002} + 3   \bchi_{020100}+ 5   \bchi_{011002}\\ &+& 4   \bchi_{101100}+ 3   \bchi_{000200} + 5   \bchi_{210010} + 5   \bchi_{100012} + 3   \bchi_{010003} +12   \bchi_{011010} \\ &+& 5   \bchi_{201001} + 8   \bchi_{002001} + 8   \bchi_{100020} + 11   \bchi_{120001} + 4   \bchi_{030000} + 20   \bchi_{100101} \\ &+& 8   \bchi_{200002} + 19   \bchi_{010011} + 3   \bchi_{310000} + 19   \bchi_{111000} + 16   \bchi_{001002} + 4   \bchi_{000003} \\ &+& 16   \bchi_{200010}+ 20   \bchi_{010100} + 26   \bchi_{001010} + 35   \bchi_{110001} + 11   \bchi_{020000}+ 4   \bchi_{300000} \\ &+& 20   \bchi_{101000} + 20   \bchi_{000011} + 23   \bchi_{000100}+ 18   \bchi_{100001} + 9   \bchi_{010000}+\bchi_{000000}\\
z_3 z_4^2&=&  \bchi_{001200} + \bchi_{012010} + \bchi_{003001} + 2   \bchi_{110110} + 2   \bchi_{101020} + \bchi_{020020} \\ &+& 3   \bchi_{000120} + 2   \bchi_{121001}+ 3   \bchi_{020101} + 4   \bchi_{101101} + 3   \bchi_{000201} + 4   \bchi_{210011} \\ &+& 3   \bchi_{201002} + \bchi_{230000} + 10   \bchi_{011011} + 4   \bchi_{112000} + 6   \bchi_{210100} + 6   \bchi_{100021} \\ &+& 4   \bchi_{002002} + 6   \bchi_{120002} + 7   \bchi_{201010} + 3   \bchi_{031000} + 12   \bchi_{011100} + 12   \bchi_{002010} \\ &+& 11   \bchi_{100102} + 16   \bchi_{120010} + 9   \bchi_{010012} + 23   \bchi_{100110} + 3   \bchi_{200003} + 6   \bchi_{310001} \\ &+& 7   \bchi_{001003} + \bchi_{000004} + 36   \bchi_{111001} + 16   \bchi_{010020} + 8   \bchi_{030001} + 16   \bchi_{220000} \\ &+& 35   \bchi_{010101} + 24   \bchi_{200011} + 40   \bchi_{001011} + 38   \bchi_{110002} + 28   \bchi_{021000} + 4   \bchi_{301000} \\ &+& 14   \bchi_{102000} + 28   \bchi_{200100} + 39   \bchi_{001100} + 13   \bchi_{300001} + 66   \bchi_{110010} + 19   \bchi_{000012} \\ &+& 58   \bchi_{101001} + 34   \bchi_{210000} + 33   \bchi_{020001} + 22   \bchi_{000020} + 58   \bchi_{000101} + 51   \bchi_{011000} \\ &+& 30   \bchi_{100002} + 51   \bchi_{100010} + 13   \bchi_{200000} + 36   \bchi_{010001} + 23   \bchi_{001000} + 8   \bchi_{000001}\\
z_3^2 z_6&=&  \bchi_{002001} + \bchi_{100101} + \bchi_{010011} + \bchi_{200002} + 2   \bchi_{111000} + 2   \bchi_{010100} \\ &+& \bchi_{001002} + 2   \bchi_{200010} + 4   \bchi_{001010} + 6   \bchi_{110001} + 3   \bchi_{020000} + \bchi_{000003} \\ &+& 4   \bchi_{000011} + \bchi_{300000} + 6   \bchi_{101000} + 6   \bchi_{000100} + 7   \bchi_{100001} + 4   \bchi_{010000} + \bchi_{000000}\\
z_3^2 z_5&=&  \bchi_{002010} + \bchi_{100110} + \bchi_{010020} + 2   \bchi_{111001} + 2   \bchi_{010101} + 2   \bchi_{200011} \\ &+& 2   \bchi_{102000} + \bchi_{220000} + 4   \bchi_{001011} + 4   \bchi_{110002} + 3   \bchi_{200100} + 3   \bchi_{021000} \\ &+& 3   \bchi_{000012} + 5   \bchi_{001100} + 10   \bchi_{110010} + 4   \bchi_{000020} + 2   \bchi_{300001} + 6   \bchi_{020001} \\ &+& 12   \bchi_{101001} + 13   \bchi_{000101} + 9   \bchi_{210000} + 9   \bchi_{100002} + 14   \bchi_{011000} + 17   \bchi_{100010} \\ &+& 6   \bchi_{200000} + 14   \bchi_{010001} + 12   \bchi_{001000} + 5   \bchi_{000001}\\
z_3^2 z_4&=&  \bchi_{002100} + \bchi_{100200} + 2   \bchi_{111010} + 2   \bchi_{010110} + \bchi_{200020} + 2   \bchi_{102001} \\ &+& \bchi_{220001} + 3   \bchi_{001020} + 3   \bchi_{200101} + 3   \bchi_{021001} + 4   \bchi_{211000} + 5   \bchi_{001101} \\ &+& 2   \bchi_{130000} + 8   \bchi_{110011} + 6   \bchi_{012000} + \bchi_{300002} + 3   \bchi_{000021} + 3   \bchi_{020002} \\ &+& 12   \bchi_{110100} + 6   \bchi_{101002} + 9   \bchi_{020010} + 3   \bchi_{300010} + 16   \bchi_{101010} + 7   \bchi_{000102} \\ &+& 3   \bchi_{100003} + 14   \bchi_{000110} + 16   \bchi_{210001} + 26   \bchi_{011001} + 24   \bchi_{100011} + \bchi_{400000} \\ &+& 11   \bchi_{201000} + 14   \bchi_{010002} + 16   \bchi_{120000} + 11   \bchi_{002000} + 29   \bchi_{100100} + 17   \bchi_{200001} \\ &+& 25   \bchi_{010010} + 28   \bchi_{001001} + 22   \bchi_{110000} + 7   \bchi_{000002} + 14   \bchi_{000010} + 6   \bchi_{100000}\\
z_3^3&=&  \bchi_{003000} + 2   \bchi_{101100} + \bchi_{000200} +    \bchi_{210010} + 3   \bchi_{011010} + 2   \bchi_{120001} \\ &+& 3   \bchi_{201001} + 3   \bchi_{002001} + 2   \bchi_{100020} + 6   \bchi_{100101} + 3   \bchi_{200002} + 4   \bchi_{010011} \\ &+& \bchi_{030000} + 2   \bchi_{310000} + 10   \bchi_{111000} + 6   \bchi_{001002} + 8   \bchi_{010100} + 9   \bchi_{200010} 
\eeqrn
\beqrn
 &+& \bchi_{000003} + 12   \bchi_{001010} + 18   \bchi_{110001} + 4   \bchi_{300000} + 10   \bchi_{000011} + 14   \bchi_{101000} \\ &+& 6   \bchi_{020000} + 14   \bchi_{000100} + 13   \bchi_{100001} + 7   \bchi_{010000} + \bchi_{000000}\\
z_2 z_4^2&=&  \bchi_{010200} + \bchi_{021010} + 2   \bchi_{001110} + 2   \bchi_{012001} + 2   \bchi_{110020} + \bchi_{003000} \\ &+& \bchi_{000030} + \bchi_{130001} + 4   \bchi_{110101} + 4   \bchi_{020011} + 4   \bchi_{101011} + 6   \bchi_{000111} \\ &+& 4   \bchi_{121000} + \bchi_{040000} + 6   \bchi_{101100} + 3   \bchi_{210002} + 7   \bchi_{011002} + 6   \bchi_{020100} \\ &+& 5   \bchi_{000200} + 6   \bchi_{100012} + 7   \bchi_{210010} + 17   \bchi_{011010} + 6   \bchi_{201001} + 16   \bchi_{120001} \\ &+& 10   \bchi_{100020} + 10   \bchi_{002001} + 24   \bchi_{100101} + 9   \bchi_{200002} + 4   \bchi_{310000} + 4   \bchi_{010003} \\ &+& 22   \bchi_{010011} + 22   \bchi_{111000} + 6   \bchi_{030000} + 17   \bchi_{001002} + 4   \bchi_{000003} + 24   \bchi_{010100} \\ &+& 17   \bchi_{200010} + 28   \bchi_{001010} + 37   \bchi_{110001} + 20   \bchi_{000011}+ 4   \bchi_{300000} + 20   \bchi_{101000} \\ &+& 12   \bchi_{020000} + 21   \bchi_{000100} + 17   \bchi_{100001} + 9   \bchi_{010000} + \bchi_{000000}\\
z_2 z_3 z_4&=&  \bchi_{011100} + \bchi_{002010} + \bchi_{120010} + \bchi_{030001} + 2   \bchi_{100110} + 3   \bchi_{111001} + 2   \bchi_{010020} \\ &+& 4   \bchi_{010101}+ 2   \bchi_{102000} + 2   \bchi_{200011} + 2   \bchi_{220000} + 5   \bchi_{001011}+ 5   \bchi_{110002} \\ &+& 3   \bchi_{000012} + 4   \bchi_{200100} + 5   \bchi_{021000} + 7   \bchi_{001100} + 13   \bchi_{110010} + 8   \bchi_{020001} \\ &+& 2   \bchi_{300001} + 5   \bchi_{000020} + 12   \bchi_{101001} + 9   \bchi_{210000} + 14   \bchi_{000101} + 8   \bchi_{100002} \\ &+& 15   \bchi_{011000} + 16   \bchi_{100010} + 5   \bchi_{200000} + 13   \bchi_{010001} + 9   \bchi_{001000} + 4   \bchi_{000001}\\
z_2 z_3^2&=&  \bchi_{012000} + \bchi_{110100} + 2   \bchi_{101010} + \bchi_{020010} + 2   \bchi_{000110} + 2   \bchi_{210001} + 4   \bchi_{011001} \\ &+& 3   \bchi_{201000} + 4   \bchi_{100011} + 4   \bchi_{120000} + 3   \bchi_{002000} + 8   \bchi_{100100} + 6   \bchi_{200001} +3   \bchi_{010002}  \\ &+& 7   \bchi_{010010} + 10   \bchi_{001001} + 10   \bchi_{110000} + 3   \bchi_{000002} + 7   \bchi_{000010} + 4   \bchi_{100000}\\
z_2^2 z_4&=&  \bchi_{020100} + \bchi_{000200} + 2   \bchi_{011010} + \bchi_{002001} + \bchi_{100020} + 2   \bchi_{120001} \\ &+& 3   \bchi_{100101} + 4   \bchi_{010011} + 4   \bchi_{111000} + \bchi_{200002} + 3   \bchi_{001002} + 2   \bchi_{030000} \\ &+& 6   \bchi_{010100} + 3   \bchi_{200010} + 7   \bchi_{001010} + \bchi_{000003} + 10   \bchi_{110001} + \bchi_{300000} \\ &+& 6   \bchi_{000011} + 5   \bchi_{020000} + 6   \bchi_{101000} + 9   \bchi_{000100} + 7   \bchi_{100001} + 4   \bchi_{010000} +\bchi_{000000}\\
z_2^2 z_3&=&  \bchi_{021000} + \bchi_{001100} + 2   \bchi_{110010} + \bchi_{000020} + 2   \bchi_{101001} + 2   \bchi_{020001} \\ &+& 3   \bchi_{000101} + 3   \bchi_{210000} +2   \bchi_{100002} + 5   \bchi_{011000} + 6   \bchi_{100010} + 3   \bchi_{200000} \\ &+& 6   \bchi_{010001} + 6   \bchi_{001000} + 3   \bchi_{000001}\\
z_2^3&=&  \bchi_{030000} + 2   \bchi_{010100} + \bchi_{001010} + 3   \bchi_{110001} + 2   \bchi_{000011} + 2   \bchi_{101000} \\ &+& 3   \bchi_{020000} + 4   \bchi_{000100} + 4   \bchi_{100001} + 5   \bchi_{010000} + \bchi_{000000}\\
z_1 z_4 z_6&=&  \bchi_{100101} + \bchi_{010011} + \bchi_{111000}  + 2   \bchi_{010100} + \bchi_{001002} + \bchi_{200010} + 3   \bchi_{001010} \\ &+& 4   \bchi_{110001} + 2   \bchi_{020000} +3   \bchi_{000011} +  3   \bchi_{101000} + 5   \bchi_{000100} + 3   \bchi_{100001} + 2   \bchi_{010000} \\
z_1 z_4 z_5&=&  \bchi_{100110} + \bchi_{010020} + \bchi_{111001} + 2   \bchi_{010101} + \bchi_{220000} + \bchi_{102000} + \bchi_{200011} \\ &+& 3   \bchi_{001011} + 3   \bchi_{110002} + \bchi_{200100} + 2   \bchi_{021000} + 2   \bchi_{000012} + 4   \bchi_{001100}+ 7   \bchi_{110010} \\ &+& 4   \bchi_{000020} + \bchi_{300001} + 5   \bchi_{020001} + 7   \bchi_{101001} + 9   \bchi_{000101} + 5   \bchi_{210000} + 6   \bchi_{100002} \\ &+& 10   \bchi_{011000} + 10   \bchi_{100010} + 4   \bchi_{200000} + 9   \bchi_{010001} + 6   \bchi_{001000} + 3   \bchi_{000001}\\
z_1 z_4^2&=&  \bchi_{100200} + \bchi_{111010} + 2   \bchi_{010110} + \bchi_{200020} + \bchi_{102001} + \bchi_{220001} \\ &+& 2   \bchi_{001020} + \bchi_{200101} + 2   \bchi_{021001} + 2   \bchi_{211000} + 4   \bchi_{001101} + 2   \bchi_{130000} \\ &+&6   \bchi_{110011} + 4   \bchi_{012000} + \bchi_{300002} + 3   \bchi_{000021} + 3   \bchi_{020002} + 8   \bchi_{110100} \\ &+&4   \bchi_{101002} + 7   \bchi_{020010} + \bchi_{300010} + 10   \bchi_{101010} + 4   \bchi_{000102} + 3   \bchi_{100003} \\ &+& 11   \bchi_{000110} + 10   \bchi_{210001} + 18   \bchi_{011001} + 16   \bchi_{100011} + \bchi_{400000} + 6   \bchi_{201000} 
\eeqrn
\beqrn
&+& 10   \bchi_{010002} + 12   \bchi_{120000} + 8   \bchi_{002000} + 17   \bchi_{100100} + 12   \bchi_{200001} + 17   \bchi_{010010} \\ &+& 17   \bchi_{001001} + 14   \bchi_{110000} + 6   \bchi_{000002} + 8   \bchi_{000010} + 5   \bchi_{100000}\\
z_1 z_3 z_6&=&  \bchi_{101001} + \bchi_{000101} + \bchi_{100002} + \bchi_{210000} + 2   \bchi_{011000} + 3   \bchi_{100010} \\ &+& 2   \bchi_{200000} + 3   \bchi_{010001} + 4   \bchi_{001000} + 2   \bchi_{000001}\\
z_1 z_3 z_5&=&  \bchi_{101010} + \bchi_{000110} + \bchi_{210001} + \bchi_{201000} + 2   \bchi_{011001} + 2   \bchi_{120000} \\ &+& 3   \bchi_{100011} + 2   \bchi_{002000} + 4   \bchi_{100100} + 4   \bchi_{200001} + 2   \bchi_{010002} + 5   \bchi_{010010} \\ &+& 7   \bchi_{001001} + 3   \bchi_{000002} + 7   \bchi_{110000} + 5   \bchi_{000010} + 4   \bchi_{100000}\\
z_1 z_3 z_4&=&  \bchi_{101100} + \bchi_{000200} + \bchi_{210010} + 2   \bchi_{011010} + \bchi_{201001} + 2   \bchi_{120001} \\ &+& 2   \bchi_{002001} + 2   \bchi_{100020} + 4   \bchi_{100101} + 2   \bchi_{200002} + 4   \bchi_{010011} + \bchi_{030000} \\ &+& \bchi_{310000} + 6   \bchi_{111000} + 3   \bchi_{001002} + 6   \bchi_{010100} + 5   \bchi_{200010} + \bchi_{000003} \\ &+& 9   \bchi_{001010} + 12   \bchi_{110001} + 2   \bchi_{300000} + 7   \bchi_{000011} + 8   \bchi_{101000} + 5   \bchi_{020000} \\ &+& 8   \bchi_{000100} + 8   \bchi_{100001} + 4   \bchi_{010000} + \bchi_{000000}\\
z_1 z_3^2&=&  \bchi_{102000} + \bchi_{200100} + 2   \bchi_{001100} + 2   \bchi_{110010} + \bchi_{000020} + \bchi_{020001} \\ &+& \bchi_{300001} + 4   \bchi_{101001} + 4   \bchi_{210000} + 3   \bchi_{000101} + 6   \bchi_{011000} + 3   \bchi_{100002} \\ &+& 7   \bchi_{100010} + 4   \bchi_{200000} + 6   \bchi_{010001} + 5   \bchi_{001000} +3   \bchi_{000001}\\
z_1 z_2 z_6&=&  \bchi_{110001} + \bchi_{020000} + \bchi_{101000} + \bchi_{000011} + 2   \bchi_{000100} + 3   \bchi_{100001} \\ &+& 3   \bchi_{010000} + \bchi_{000000}\\
z_1 z_2 z_5&=&  \bchi_{110010} + \bchi_{020001} + \bchi_{101001} + \bchi_{000020} + 2   \bchi_{000101} + \bchi_{210000} + 2   \bchi_{100002} \\ &+& 3   \bchi_{011000}+ 4   \bchi_{100010} + 2   \bchi_{200000} + 5   \bchi_{010001} + 4   \bchi_{001000} + 3   \bchi_{000001}\\
z_1 z_2 z_4&=&  \bchi_{110100} + \bchi_{101010} + \bchi_{020010} + 2   \bchi_{000110} + \bchi_{210001} + 3   \bchi_{011001} \\ &+& \bchi_{201000} + 3   \bchi_{120000} + 3   \bchi_{100011} + 2   \bchi_{002000} + 5   \bchi_{100100} + 3   \bchi_{200001} \\ &+& 2   \bchi_{010002} + 6   \bchi_{010010} + 6   \bchi_{001001} + 6   \bchi_{110000} + 2   \bchi_{000002} \\ &+& 4   \bchi_{000010} + 2   \bchi_{100000}\\
z_1 z_2 z_3&=&  \bchi_{111000} + \bchi_{010100} + \bchi_{200010} + 2   \bchi_{001010} + 3   \bchi_{110001} + 2   \bchi_{000011} + 2   \bchi_{020000} \\ &+& \bchi_{300000}+ 4   \bchi_{101000} + 4   \bchi_{000100} + 5   \bchi_{100001} + 3   \bchi_{010000} + \bchi_{000000}\\
z_1 z_2^2&=&  \bchi_{120000} + \bchi_{100100} + 2   \bchi_{010010} + \bchi_{200001} + 2   \bchi_{001001} + \bchi_{000002} \\ &+& 4   \bchi_{110000} + 3   \bchi_{000010} + 3   \bchi_{100000}\\
z_1^2 z_6&=&  \bchi_{200001} + \bchi_{001001} + 2   \bchi_{110000} + \bchi_{000002} + 2   \bchi_{000010} + 3   \bchi_{100000}\\
z_1^2 z_5&=&  \bchi_{200010} + \bchi_{001010} + 2   \bchi_{110001}  + 2   \bchi_{000011} + \bchi_{020000} + 2   \bchi_{101000} \\ &+& 3   \bchi_{000100} +4   \bchi_{100001} + 3   \bchi_{010000} + \bchi_{000000}\\
z_1^2 z_4&=&  \bchi_{200100} + \bchi_{001100} + 2   \bchi_{110010} + \bchi_{000020} + \bchi_{020001} + 2   \bchi_{101001} \\ &+& 3   \bchi_{000101} + 2   \bchi_{210000} + 4   \bchi_{011000} + \bchi_{100002} + 5   \bchi_{100010} + \bchi_{200000} \\ &+& 4   \bchi_{010001} + 3   \bchi_{001000} + \bchi_{000001}\\
z_1^2 z_3&=&  \bchi_{201000} + \bchi_{002000} + 2   \bchi_{100100} + 2   \bchi_{200001} + \bchi_{010010} + 3   \bchi_{001001} \\ &+& \bchi_{000002} + 4   \bchi_{110000} + 3   \bchi_{000010} + 2   \bchi_{100000}\\
z_1^2 z_2&=&  \bchi_{210000} + \bchi_{011000} + 2   \bchi_{100010} + 2   \bchi_{200000} + 2   \bchi_{010001} + 3   \bchi_{001000} + 2   \bchi_{000001}\\
z_1^3&=&  \bchi_{300000} + 2   \bchi_{101000} + \bchi_{000100} + 3   \bchi_{100001} + 2   \bchi_{010000} + \bchi_{000000}
\end{eqnarray*}

\label{lastpage}

\end{document}